\documentclass[openacc]{rstransa}

\usepackage[utf8]{inputenc}
\usepackage[normalem]{ulem}
\usepackage{soul}

\newcommand{\apj}{ApJ}
\newcommand{\apjs}{ApJS}

\newcommand{\aap}{A{\&}A}

\newcommand{\mnras}{MNRAS}

\newcommand{\solphys}{{\it Solar Phys.}}
\newcommand{\ssr}{{\it Space Sci. Rev.}}
\newcommand{\jgr}{{\it J. Geophys. Res.}}
\newcommand{\beq}{\begin{equation}}
\newcommand{\eeq}{\end{equation}}
\newcommand{\bea}{\begin{eqnarray}}
\newcommand{\eea}{\end{eqnarray}}
\newcommand{\p}{\partial}
\newcommand{\mb}{\mathbf}

\newcommand*{\doi}[1]{\href{http://dx.doi.org/#1}{doi: #1}}

\jname{rsta}
\Journal{Phil. Trans. R. Soc}

\begin{document}

\title{Sources of Solar Energetic Particles}

\author{Loukas Vlahos$^{1}$, Anastasios Anastasiadis$^{2}$, Athanasios Papaioannou$^{2}$, Athanasios Kouloumvakos$^3$  and Heinz Isliker$^{1}$ }

\address{$^{1}$ Department of Physics,
              Aristotle University, \\54124 Thessaloniki, Greece\\
$^{2}$ Institute for Astronomy, Astrophysics, \\Space Applications and Remote Sensing,\\
National Observatory of Athens\\
GR-15236 Penteli, Greece

$^{3}$IRAP, Universit\'e de Toulouse III - Paul Sabatier,
CNRS, CNES, UPS, Toulouse, France
}

\subject{xxxxx, xxxxx, xxxx}


\corres{Loukas Vlahos\\
\email{vlahos@astro.auth.gr}}

\begin{abstract}

Solar Energetic Particles (SEP) are an integral part  of the physical processes related with Space Weather. We present a review for the acceleration mechanisms related to the explosive phenomena (flares and/or  CMEs) inside the solar corona. For more than 40 years, the main 2D cartoon representing our understanding of the explosive phenomena inside the solar corona remained almost unchanged.  The acceleration mechanisms related to solar flares and CMEs also remained unchanged and were part of the same cartoon.  
In this review, we revise the standard cartoon and present evidence from  recent global  MHD simulations that supports the argument that explosive phenomena will lead to   the spontaneous formation of current sheets in different parts of the erupting magnetic structure. The evolution of the large scale current sheets and their fragmentation  will lead to strong turbulence and turbulent reconnection during solar flares  and turbulent shocks. In other words, the acceleration mechanism in flares and CME-driven shocks may be the same, and their difference will be the overall magnetic topology,  the ambient plasma parameters, and the duration of the unstable driver.

\end{abstract}

\maketitle

\section{Introduction}\label{Intro}
Solar Energetic Particles (SEP) contain important information about the mechanisms of the particle energization inside the solar corona, as well as the properties of the acceleration volume. So far two classes of SEP events are observed and are classified as ``gradual'' and as ``impulsive''. The two classes have several distinct characteristics. The impulsive events are associated with short time scales, large electron to proton and $^3$He/$^4$He ratios, and high ionisation state, indicating a source region with temperature $~ 3\times 10^7$ K. Gradual events are associated with larger time scales and coronal or inter planetary shocks, high proton intensities, energetic abundances similar to the solar corona, and charge states corresponding to the source region with temperature $~ 3 \times 10^6$ K \cite{Reames95}. Classifications are always useful in order to divide events originating from different physical processes.  However, observations from Advance Composition Explorer (ACE) spacecraft revealed rich structure in the time dependence of the two classes of SEP. A recent analysis of ACE data indicates that many events defy a simple classification as ``impulsive'' or ``gradual''. Gradual events possess an impulsive part, suggesting a clear synergy between flare accelerated particles with shock acceleration. On the other hand, gradual events related with radio, X-ray and gamma-ray observations indicate that particle acceleration takes place in large scale coronal structures behind the CME, and that the CME is not the sole acceleration region in gradual events \cite{Klein01}.

\begin{figure}[h!]
\centering
\includegraphics[width=0.7\textwidth]{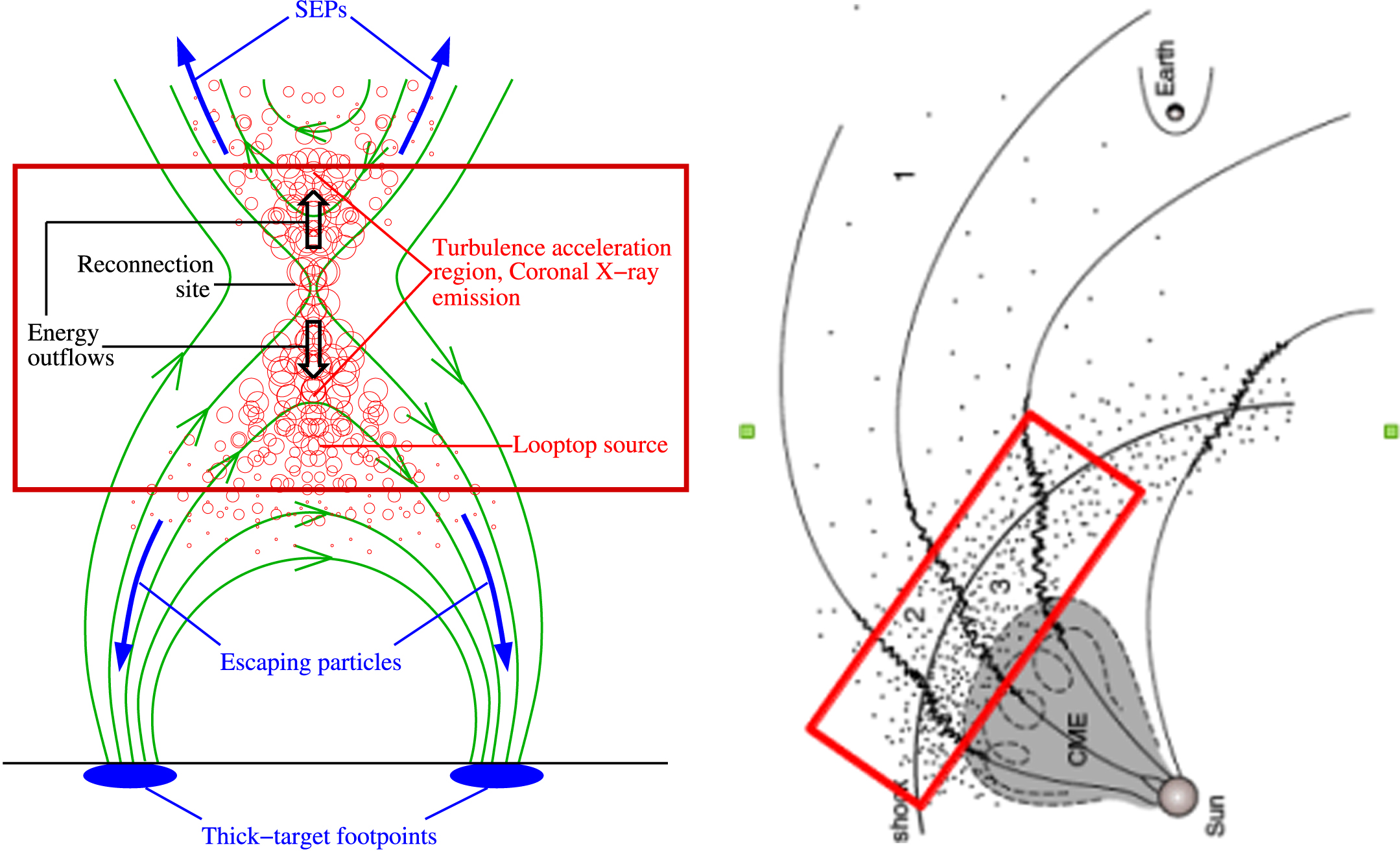}
\caption{Left hand panel: schematic representation of the reconnecting field (solid green) forming closed coronal loops and open field lines, presumably extending higher up into the corona and the solar wind. The red foam represents turbulence. Acceleration probably takes place in the outflow regions above and below the X-point. Particles (temporarily) trapped here produce the radiation seen above the closed loops, and particles escaping these regions up and down (blue arrows) are  observed at 1 AU as SEPs and produce the non-thermal radiation (mainly at the two footpoints; blue ovals), respectively. Right hand panel: similar schematic view, joining the flare site field lines to the CME, the shock, and beyond (from Lee 2005, \cite{Lee2005}). The rectangles define the boundary of the acceleration sites and represent the leaky box.\cite{Petrosian16}\label{petro}}
\end{figure}

A theory of particle acceleration in solar flares must explain how electrons and ions are energized out of thermal plasma, as well as, should provide time scales, energy spectra, fluxes, and abundance ratios for various particle species. At the same time it should address the characteristics of the induced emission of radio waves, X-rays, gamma-rays and neutrons. The current stage of the theory of particle acceleration during solar eruptions is based on a static 2D picture of the so called  ``standard'' cartoon for solar flares, see Fig.\ \ref{petro}. In this picture,  the acceleration of electrons is based on the reconnection at the current sheet formed below the large scale CME magnetic structure \cite{Petrosian16}. The stochastic acceleration of particles  by the weak turbulence driven by the jets of the reconnecting current sheet has been analysed in Ref.\ \cite{Petrosian12}.

Alternatives to the above scenario are relying on the evolution and  fragmentation of the monolithic current sheet and the re-acceleration of particles in many current sheets formed along the erupting magnetic structure \cite{Cargill12}. The stochastic acceleration of ions by weak MHD turbulence serves as the main mechanism for the  abundance enrichment mentioned earlier \cite{Reames1999,Petrosian12}. Unfortunately this mechanism cannot account for the efficient acceleration of $^3$He and  alternative ideas have been proposed in the literature \cite{Miller97, Aschwanden02}.  The ``standard'' cartoon for solar flares cannot explain the fast escape for the impulsive SEP electrons and ions. Its purpose was to model mainly the radio and X-ray emission from low coronal loops. At the same time, the CME driven shocks can also efficiently accelerate ions (see the reviews \cite{Krauss-Varban10, Lee2012, Desai16}).

In this review, we stress two important points: (a) How the photospheric turbulence drives the {spontaneous formation of current sheets during solar eruptions (see  
the brief  discussion in section 3, where we emphasise the data driven MHD modelling, and see more details in the article by Archontis and Syntelis \cite{Archontis19} in this issue), and (b) how the unstable large scale magnetic topologies sustain strong turbulence, which is crucial for particle acceleration in solar flares and CME-driven shocks. We believe that current data-driven large scale 3D MHD simulations are closer to a realistic scenario for the acceleration of the SEP \cite{Inoue16}.

This review is organised as follows: In section 2 we outline the main observational constraints on current theories on SEP acceleration theories. In section 3 we discuss the evolution of the data driven large scale magnetic topologies during an eruption. In section 4 we point out the difference between weak and strong turbulence, and in section 5 we analyse the acceleration of particles during magnetic flux emergence. In section 6 we briefly discuss the mechanisms for particle acceleration in CME driven shocks, and in section 7 we outline the main points of this review.

\section{Key observational constraints for particle acceleration theories in the Solar Corona}\label{Obs}

\subsection{Location of SEP acceleration and release sites}

The detection of an SEP event at any given location in the heliosphere requires a direct magnetic connection to the region where the particle acceleration and release takes place. When the accelerated charged particles are released into open magnetic field lines, they propagate from their production site through the interplanetary medium, spiraling along the field lines until their detection through instruments that perform in-situ measurements. Since particles tend to move more easily along the magnetic field lines than across, the relative location of the parent active region (AR) with respect to the location of the observing point is considered to be an important parameter for the detection of an SEP event.

Eruptions that takes place at ARs with western locations on the solar disk, seem to be more probable to record a SEP event, due to the Parker spiral shape of the interplanetary magnetic field \cite{Belov2005}. This picture is mostly consistent with the impulsive-flare-related SEP events being associated with localized sources close to the Sun. Gradual-CME-related SEP events are usually detected at widely separated locations in the heliosphere \cite{Anastasiadis18,Richardson2014}. Remote sensing observations and in-situ measurements from multiple vantage points (SoHO, STEREO, ACE, and Wind) have significantly improved our knowledge about the large longitudinal spread of SEP events and also helped to define the locations of the possible particle acceleration and release sites. Widespread (i.e. $>$90--180$^\circ$) and usually long duration (i.e. $>$2 hours) SEP events (Gradual-CME-related SEPs), are associated with the spatial and temporal evolution of coronal and interplanetary shock waves that are presumably driven by fast and wide CMEs. Additionally multi-spacecraft observations have challenged in some cases the expectation of a narrow longitudinal distribution of impulsive SEP events. The analysis of Wiedenbeck et al. \cite{Wiedenbeck2013} showed that impulsive events can also be detected over wide longitudinal ranges.

A combination of several data sets, including remote-sensing and in-situ observations, made possible to relate the different aspects of the eruptive events to the SEP events seen at 1 AU. For the gradual-CME-related SEP events, the CME-driven shock waves are considered to have an important role in the acceleration and the release of SEPs over a wide range of longitudes. Rouillard et al. \cite{Rouillard2012}, by using multi-viewpoint observations of the CME and the solar corona for an eruptive event that occurred on 21 March 2011, showed that the CME extension over a wide longitudinal range in the corona was responsible for the large longitudinal spread of an SEP event. A long list of continued studies signified even further the role of the expanding shock waves \cite{Lario2014,Kouloumvakos2016,Lario2016,Lario2017,Kwon2018} or EUV waves \cite{Park2013,Prise2014,Park2015,Miteva2014} in the longitudinal spread of SEP events as observed at distant magnetically connected s/c. In any case, there is a converging argument that the acceleration process itself occurs over a wide range of longitudes rather than in a small source region, where the accelerated particles could be transported before being injected at distant longitudes.

\subsection{Critical properties of CMEs and the associated shock waves}

A series of studies have shown a significant correlation of SEP peak intensities with the speed and other properties of the associated CMEs \cite{Reames1999,Kahler2000, Kahler2001,Kahler2005,Kahler2013,Richardson2015,Papaioannou2016}. Kahler et al. \cite{Kahler2001} showed that the speed of the CMEs associated with SEPs have a good correlation with the energetic proton peak intensities, and additionally Kahler et al. \cite{Kahler2013} found that the total CME energy is well correlated with either the SEP peak intensities or the total SEPs energy. Furthermore, Richardson et al. \cite{Richardson2015} compared the estimates of CME parameters using several catalogs to reduce the projection effects for the correlations with the proton intensities. They found that the CME speed in quadrature when compared with the SEPs peak intensity results in higher correlations.

The comparisons between critical parameters of CMEs and SEP properties have revealed significant correlations, however, in any case there is a large spread ($\sim$3 decades) in the SEP properties for a given CME parameter. One of the important factors for the large spread in the correlations may arise from the use of CME parameters as a proxy for the CME-driven shock wave characteristics. In addition, the involved uncertainties in determining the CME speed and other parameters, while they are projected measures in the plane of the sky, may also be an important factor. Nevertheless the CME parameters, such as their speed and width, can be easily inferred from remote sensing observations and serve as a rather good basis of the current SEP forecasting schemes \cite{Anastasiadis18}.

The role of shock waves in energetic particle acceleration and release has been further elucidated since the constantly evolving techniques that have been developed, made possible to infer shock parameters from remote-sensing observations. Those techniques integrate the observations provided by instruments located at multiple viewpoints. The density compression ratio and Alfv\'enic Mach number are some of the critical shock parameters that can be deduced from remote-sensing observations \cite{Ontiveros2009, Bemporad2014,Kouloumvakos2014,Kwon2018}. Kwon et al. \cite{Kwon2018} showed that there is a wide spatial range where the shock waves, analyzed in their study, are super-critical for a long duration with an average density compression ratio of 2.1--2.6. This result provides additional support to the expectation of previous studies that the wide extent of these shocks is the reason for the distribution of SEPs over a very wide range of heliospheric longitudes.

Additionally, MHD shock modeling \cite{Kozarev2013,Kozarev2015} or a combination of shock forward modeling with parameters of the background corona from MHD simulations \cite{Rouillard2016}, seem to be essential to determine the shock wave properties at the apparent acceleration/release sites of SEPs. The studies of Kozarev et al. \cite{Kozarev2013,Kozarev2015} and Rouillard et al. \cite{Rouillard2016} suggest a link between the shock characteristics and the SEPs measured in-situ (see also \cite{Afanasiev2018}). Additionally, Kouloumvakos et al. (2019), from an extensive comparison between the 3-D shock parameters at the magnetically well-connected regions and the SEPs characteristics, showed significant correlations, better than those with the CME properties. The best correlation is established for the comparison between the shock Alfv\'enic Mach number and the SEPs peak intensity $>70\%$. Additionally, Kouloumvakos et al. (2019) showed that the evolution of the shock parameters at the well connected regions can sufficiently explain the characteristics of the observed SEP events even for the far connected cases. These results signify further the role of the shock waves in particle acceleration and additionally seem to establish a very good association between the longitudinal extent of the SEP event in the heliosphere and that of the lateral extension of the CME and the CME-driven shock wave.

\subsection{Impulsive, Gradual and Mixed events}

The dichotomous picture of SEP events has proved to be useful, however, both flares and CMEs are associated with almost all kind of SEP events, so it is often difficult to distinguish the particle accelerator unambiguously. Additionally, there are SEPs with properties that fail to strictly follow the dichotomous picture \cite{Cane2010}. This indicates that the classical picture might be a simplification and a third category of events could exist \cite{Papaioannou2016}. This third category, so-called \textit{hybrid or mix events}, have SEP properties that resemble gradual SEP events, but also demonstrate properties of impulsive SEPs. In this category the SEP events may result from both solar flare or CME-driven shock acceleration. In this context the acceleration mechanism should be a combination of stochastic acceleration and shock drift acceleration in different timescales of the SEP evolution.

The difficulty to make a dichotomous separation of SEP events is also reflected in studies that examine the association between type III, II, and IV radio bursts and SEP events. Radio emissions have a rich diagnostic potential about the acceleration and propagation of solar energetic particles as well as shocks. Miteva et al. \cite{Miteva2013} showed that SEP events, either gradual or impulsive, were found to have the highest association rate with type III radio bursts and a lower association with type II bursts. Additionally, Kouloumvakos et al. \cite{Kouloumvakos2015} examined the association of the SEP release time, as inferred by the velocity dispersion analysis, with transient solar radio emissions to identify the most relevant acceleration processes. Their study showed that both flare- and shock-related particle release processes are observed in major proton events at $>$50 MeV and a clear-cut distinction between flare-related and CME-related SEP events is difficult to establish. The proton release was found to be most often accompanied by both type III and II radio bursts, but also a good association rate only with type III radio bursts was found.

}

\subsection{Elemental abundance and spectral variability}

The abundance of elements and isotopes of SEPs have been extensively used as indicators of their acceleration and transport and also served as the earliest indication of the dichotomous picture of SEPs \cite{Reames2018a}. Gradual events are in general considered to have a composition similar to that of the corona or solar wind, while impulsive events typically have enhanced element and isotope ratios.

\begin{figure}[h!]
\centering
\includegraphics[width=0.7\textwidth]{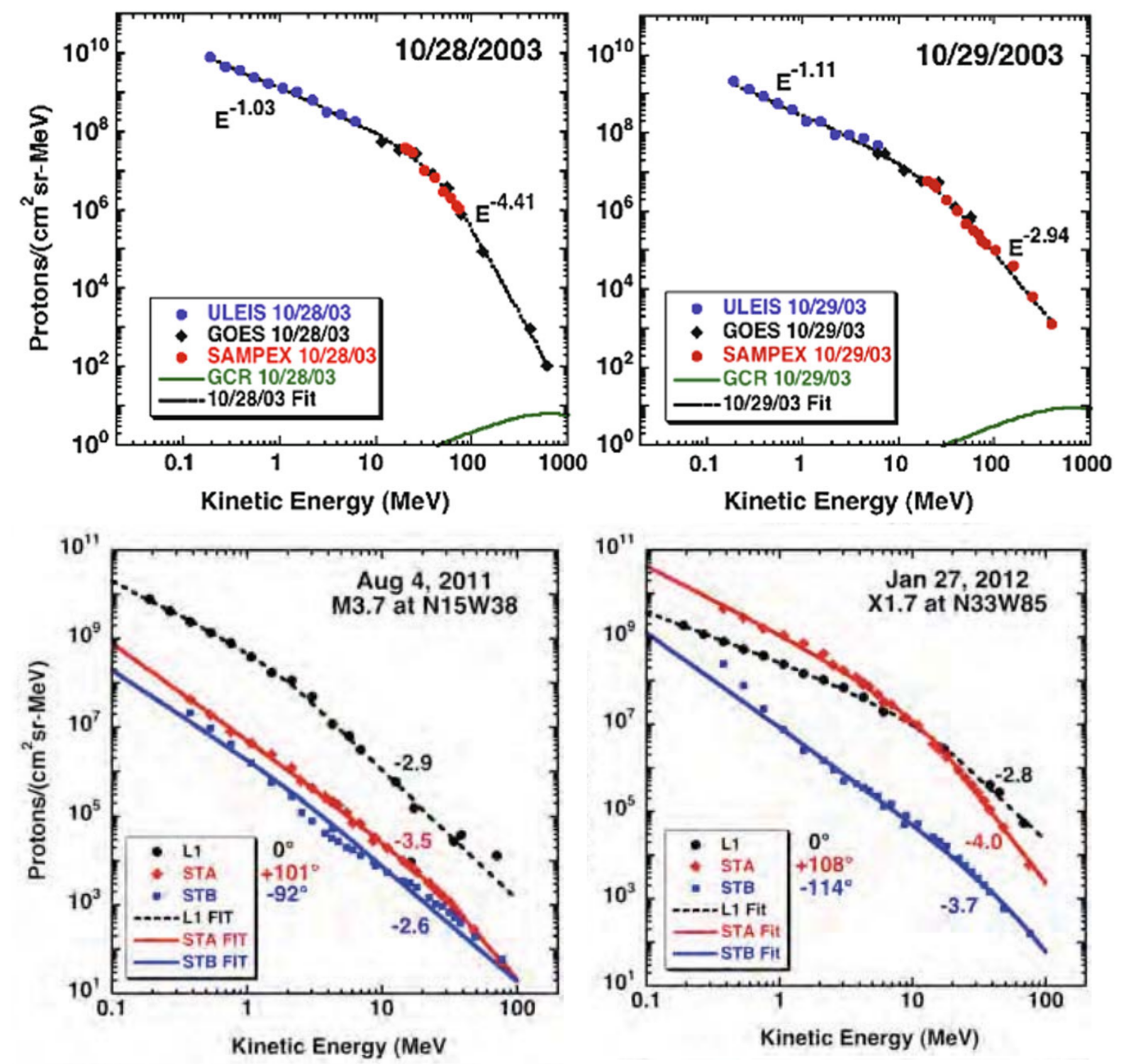}
\caption{Top panels: Proton fluence spectra for two GLE events from Mewaldt et al. (2012).
 Bottom panels: Proton fluence spectra from multy-spacecraft observations (STA, STB, and near-Earth)
  for two large SEP events, from Mewaldt et al. (2013). (Left: 4 August 2011; Right: 27 January 2012).
   The measured spectra are fitted with a double power-law function.}
\label{fig:spectra}
\end{figure}

The SEPs energy spectra and their variability is another important characteristic that provides useful information about the acceleration processes involved. From an analysis of 16 Ground Level Enhancements (GLE), Mewaldt et al. (2012) found that the energetic proton spectra exhibited breaks between $\sim$2 and 50~MeV and that they were better represented by a double power-law function (see Figure~\ref{fig:spectra}). This study also showed that GLE spectra are harder, with spectral indices $\gamma \sim$3 above 40~MeV/nucleon, in comparison with other SEP events. Evidence of spectral hardening might suggest that a different or a more complex acceleration process could dominate at higher energies \cite{Anastasiadis18}. Additionally, Mewaldt et al. (2013) showed that for some large multi-spacecraft SEP events there is a wide divergence in spectral slopes for the same event. The energy spectrum of the SEPs that were observed by the best connected spacecraft to the source region exhibits an energy spectrum above the spectral break as hard or harder than the others, with the exception of the 3 Nov. 2011 event. The spectral differences may be attributed to different shock geometry, the relative contribution of shock acceleration and downstream turbulent reconnection \cite{Zank15, leRoux16, Afanasiev14, Garrel18}, or the acceleration in the current fragmentation of the eruptive large scale magnetic structures.

\subsection{What about SEP without Flares or CMEs?}

Flares and coronal mass ejections (CMEs) are the two main manifestations of solar eruptions and typically they accompany each other. However, a significant fraction of intense flares seem not to be accompanied by CMEs \cite{Wang2007}. Those cases have been termed as confined flares since they are unaccompanied with any ejection signature. Some of those confined events were additionally lacking any SEP signature, even of those that were located at the western solar hemisphere \cite{Klein2010}. Klein et al. \cite{Klein2011}, concluded that a possible reason why major solar flares in the western hemisphere are not associated with SEPs is the confinement of particles accelerated in the impulsive phase. On the other hand, only a few SEP events are associated with a CME and no flare signatures in the corona. An old paradigm is an event presented in Kahler et al. \cite{Kahler1986}, which included a filament eruption and a CME without an impulsive flare or radio emission in the low corona, and it was associated with an SEP event.

\subsection{A summary of the key observational points}
A complete theory of the coronal sources of the  Solar Energetic Particles is currently absent since many key observations are still missing.  A few important points discussed in this section are:
\begin{enumerate}
	\item The impulsive-flare related SEP events are associated with localized sources close to the Sun. Gradual-CME related SEP events are usually determined at widely separated locations in the heliosphere.
	\item SEP events with characteristics resembling impulsive events can also be detected over wide longitudinal ranges.
	\item  The simplistic dichotomy of the SEP events in impulsive and gradual is not present in all SEP events.		
	\item Observations are consistent with the acceleration process occurring  over a wide range of longitudes rather a small source region, and the accelerated particles could be transported before being injected at distant longitudes.
	\item The comparison between CME or shock parameters and SEP properties shows significant correlations, better than the correlations with flare parameters.

	\item SEP events, either gradual or impulsive, were found to have high association with both Type III and Type II radio bursts.
	
	\item Gradual events are in generally  considered  to have a composition similar to that of the corona or the solar wind, while impulsive events typically have enhanced element and isotope ratios.
	\item The energy spectra based on the Ground Level Enhancements show a double power-law, with the break between 2-50 MeV.
	\item Observations have revealed that most of the SEP events are associated with both flares and CMEs. Several intense flares that seem not to be accompanied by a CME also were lacking SEPs.
	\end{enumerate}
We will discuss  in section 7 how the key observational points listed above can be interpreted  by the  acceleration mechanisms presented in this review.

\section{The evolution of magnetic topologies and eruptive phenomena}

Solar ARs, coronal holes and the quiet sun are all driven by the turbulent photospheric flows and the emergence of new magnetic flux (see Archontis and Syntelis \cite{Archontis19} in the current issue).

\begin{figure}[h!]
	\centering
		\includegraphics[width=0.4\textwidth]{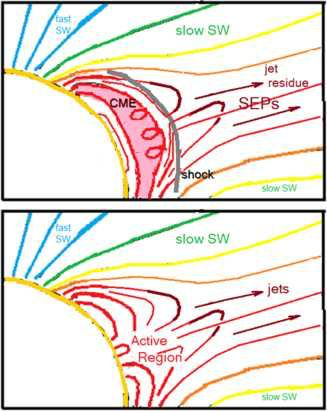}

		\caption{Sketch of possible sources of SEPs. The upper panel shows a CME-driven shock wave (gray)  that accelerates plasma from the high corona, and residues from jets. Blue field lines track the fast solar wind (SW) from coronal holes that have photoshperic sources similar to the slow solar wind, but less trapping and divergence. The lower panel shows an active region (red), containing closed loops  from which solar jets emerge. Field lines carrying the slow solar wind (yellow and green) diverge from open field lines from the photosphere outside of active regions.  \cite{Reames2018a} }
	\label{topologies}
\end{figure}
The acceleration mechanisms for SEP are closely related  with the evolution of the 3D eruptive magnetic topologies and the energy release processes \cite{Reames2018a}. The magnetic eruptions relate  all the known  particle acceleration mechanisms (strong turbulence and CME related shocks). In Fig. \ref{topologies} a sketch of the magnetic field topologies closely related with the observed properties of SEP events is presented.  It is important to connect the sketch in Fig.\ \ref{topologies}   with the evolution of eruptive magnetic topologies driven by turbulent photospheric motions\cite{Inoue16}. In the following, we present examples of eruptive  magnetic topologies using as initial topology Non-Linear Force Free extrapolations of observed magnetograms, as driven by photospheric motions \cite{Kliem13, Inoue18, Jiang18}.

\begin{figure}[h!]
	\centering
		\includegraphics[width=0.6\textwidth]{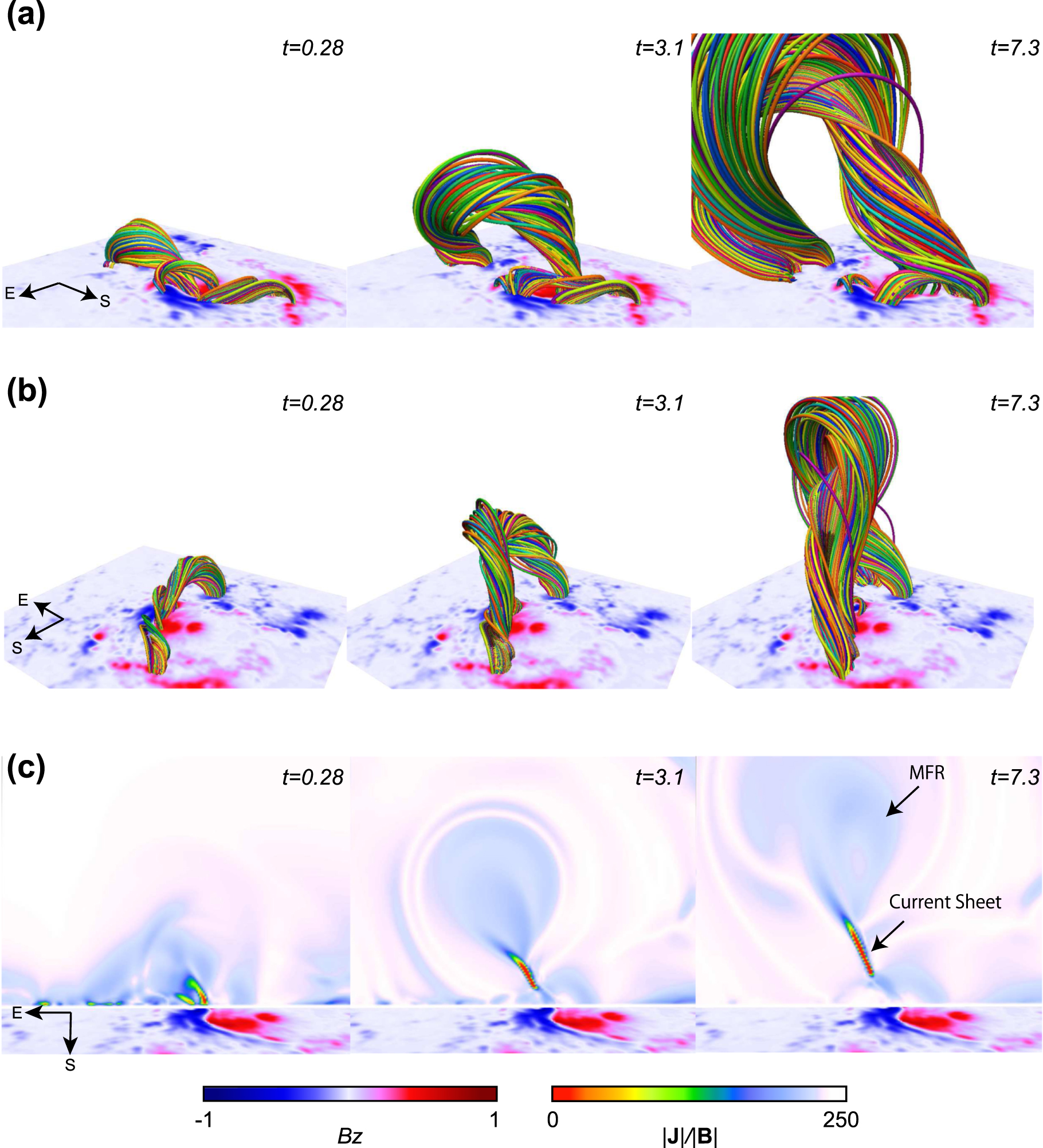}

			\caption{Temporal evolution of the formation and dynamics of an eruptive magnetic flux rope (MFR). Panels (a) and (b) show the field lines from different viewing angles. E and S stand for east and south. An animation of these panels is available in the original article, its duration is 4 s. (c) Temporal evolution of |J|/|B| 
			plotted in the x-z plane at y = 0.38, with a viewpoint from the south.\cite{Inoue18}}
	\label{inoue1}
	\end{figure}
 Inoue et al. \cite{Inoue18}  performed a magnetohydrodynamic
(MHD)
simulation in order to reveal the three-dimensional
dynamics of the magnetic
fields associated with a X9.3 solar
flare. They first
performed an extrapolation of the 3D magnetic
field based on the observed photospheric magnetic
field prior to the
flare and then used this as the initial condition for the MHD simulation, which revealed a dramatic eruption, see Fig.\ \ref{inoue1}. In
particular, they found that a large coherent
flux rope composed of highly twisted magnetic
field lines formed during
the eruption. A series of small
flux ropes were found to lie along a magnetic polarity inversion line prior to the
flare. Reconnection occurring between each
flux rope during the early stages of the eruption formed the large,
highly twisted
flux rope. Furthermore, they observed a writhing motion of the erupting
flux rope. Understanding
these dynamics is of high significance  in order to increase the accuracy of space weather forecasting. Inoue et al. \cite{Inoue18} reported on the
detailed dynamics of the 3D eruptive
flux rope and discussed the possible mechanisms of the writhing motion. The brief presence of  a large scale current sheet is apparent but soon after it fragments, forming  a turbulent environment in the same location.

\begin{figure}[h!]
	\centering
		\includegraphics[width=0.7\textwidth]{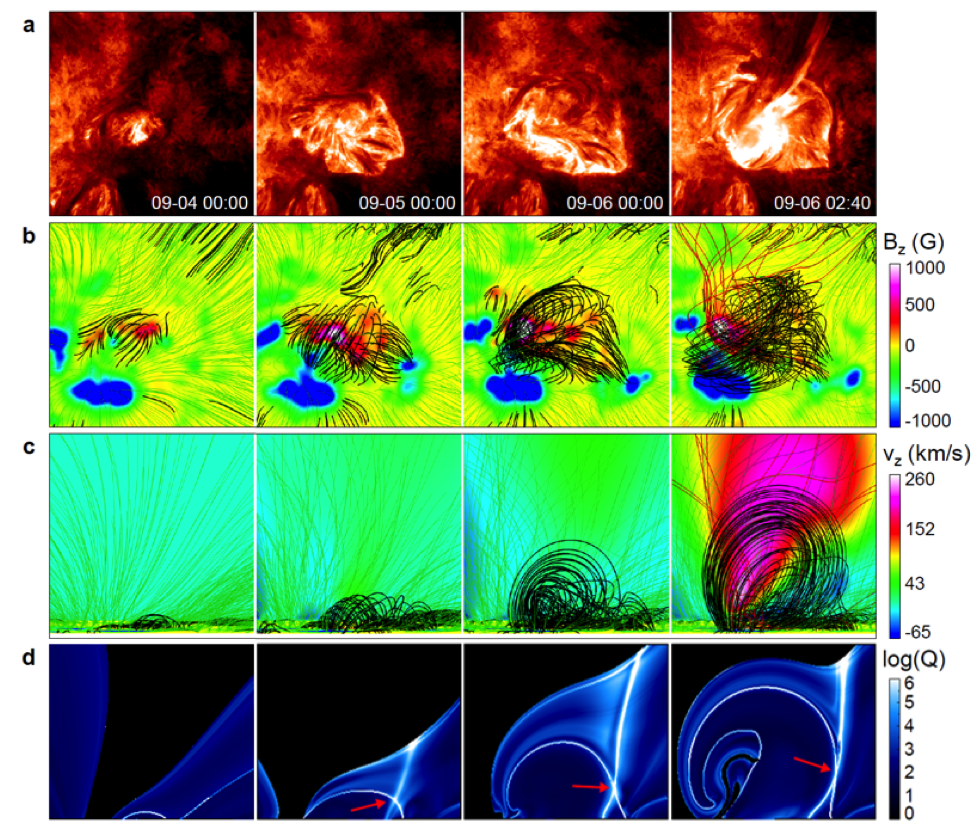}

			\caption{(a) Images at different times from the initial emergence to the eruption. (b) Top view of the corresponding magnetic field evolution at different times (t=0, 12, 24, 48 and 57) from the MHD model. The field lines are traced from footpoints evenly distributed at the bottom surface, which is shown with the photospheric magnetic flux map. Field lines closed (open) in the box are colored black (green), while those becoming open from closed during the eruption are colored red. (c) Side view of the magnetic field lines from south (that is, the horizontal and vertical axes are x and z, respectively). The background shows a 2D central cross-section of the 3D volume and its color indicates the value of the vertical component of the velocity. (d) Vertical cross sections of the evolving magnetic topology show the spontaneous formation of a large scale current sheet. \cite{Jiang18}  }
	\label{topologies1}
	\end{figure}
	 In Fig. \ref{topologies1}, 3D simulations of a data driven eruption or flux emergence,
	 interacting with the ambient magnetic field and forming a blow out jet, is shown \cite{Jiang18}. The MHD model of Jiang et al. \cite{Jiang18} was initialized with a potential field extrapolation of the vertical component of the photospheric field, and a highly tenuous plasma in hydrostatic, isothermal state (with solar gravity) was assumed to approximate the coronal low-$\beta $ plasma condition.
	 	 They drive the model continuously by supplying the bottom boundary with a data stream of photospheric vector magnetograms.
	 The Heliospheric Magnetic Imager (HMI) provides routinely high-quality vector magnetograph data at the photosphere with
	 spatial resolution of  $1 arcsec$ and cadence  of $12 m$,
	 which is adequate for tracking the relatively long-term (hours to days) evolution of AR magnetic structures from their formation to eruption.
	 To ensure that the input of boundary vector fields is self-consistent, they utilize the method of projected characteristics, which has its foundation in the wave-decomposition principle of the full MHD system.
	 It has been shown that such a method can naturally simulate the transport of magnetic energy and helicity to the corona from below.

	In the simulations reported  above,  the spontaneous formation of reconnecting current sheets at several locations in the evolving structures is apparent. On the other hand,
	formed large scale reconnecting current sheets are also fragmented and disappear rapidly, forming a turbulent reconnecting environment \cite{Onofri06,Cargill12,Dahlin15}.  In realistic magnetic topologies, being led to eruption through photospheric turbulent flows or the emergence of new magnetic flux, 
	distributed reconnecting current sheets and large amplitude magnetic disturbances will always be present. This is in contrast to the static 2D cartoon (Fig.\ \ref{petro})  used extensively in the current literature and referred to as the ``standard'' flare model, where the monolithic current sheet is artificially maintained for long times and its jets generate weak turbulence \cite{Petrosian16}.

	Numerous articles that provide examples of MHD simulations of erupting flux ropes leading to the formation of current sheets have been reported in the past. These models initiate the simulations with an artificial  magnetic loop stressed by simple photospheric motions. The main goal of these simulations was to reproduce the standard flare model shown in Fig. \ref{petro}.

	\section{Energisation of particles in weak and strong turbulence during explosive events}

		\subsection{Weak turbulence}

	In the astrophysics community the term ``turbulence'' is synonymous to ``weak turbulence'' and refers to	 stochastic interaction of particles with low amplitude $(\delta B/B<<1)$ linear  MHD waves \cite{Melrose2009, Miller97}  .
	The stochastic  (or second order) acceleration of particles was first proposed and analysed by  Fermi \cite{Fermi49} as a mechanism for the acceleration of Cosmic Rays \cite{Longair11}. The core of his idea  had a larger impact on non-linear processes in general and has been the driving force behind all subsequent theories on charged particle energization in space and astrophysical plasmas. In the original treatment, relativistic particles were accelerated by collisions with very massive, slowly moving magnetic clouds (scattering centers).  The rate of the systematic energy gain of the charged particles with the scatterers is proportional to the square of the ratio of the magnetic cloud speed ($V$) to the speed of light ($c$), i.e.~$(V/c)^2$. A few years after the initial article by Fermi, Davis \cite{Davis} and Parker and Tidman \cite{Parker58} emphasized the stochastic nature of the initial Fermi proposal and they estimated analytically the transport coefficients, using an idealized assumption for the interaction of the scatterers with the particles. Parker and Tidman \cite{Parker58} assumed that the scattering centers are randomly moving and applied their model to solar flares, accelerating protons from the thermal distribution.

The initial idea put forward by Fermi was soon replaced in the astrophysical literature with a new suggestion based on the interaction of charged particles with a Kolmogorov spectrum of low amplitude MHD waves $(\delta B/B<< 1)$,  and the acceleration process was renamed as stochastic (weak) turbulent heating and acceleration or simply stochastic acceleration by turbulence (\cite{Davis, Tverskoi67, Kulsrud71};  see also the reviews by Miller and Petrosian \cite{Miller97, Petrosian12}). When the amplitude of the waves $(\delta B)$ is much smaller than the mean magnetic field $B$, the transport coefficients  are estimated with the use of the quasilinear approximation, and by solving the transport equations one can estimate the evolution of the energy distribution of the particles \cite{Achterberg81,Schlickeiser89}. The Fokker-Planck equation became the main tool for the analysis of the evolution of energy distributions of particles. The diffusion (Fokker-Planck) equation estimates the rate of change of the energy distribution $n(W,t)$ of the accelerated particles. In order to simplify the diffusion equation, spatial diffusion was dropped, since it was assumed that it is not important  inside the acceleration region,
        \begin{equation} \label{diff}
            \frac{\partial n}{\partial t} +
            \frac{\partial}{\partial W} \left[F n -\frac{\partial (D n) }{\partial W} \right] =
            -\frac{n}{t_{\rm esc}} + Q ,
        \end{equation}
    where $t_{\rm esc}$ is the escape time from an acceleration volume with characteristic length $L$, $Q$ is the injection rate, $F$ and  $D$ are the transport coefficients. The escape time is also kept as a free parameter in most applications of stochastic acceleration.

The main disadvantage of the stochastic acceleration of particles through low amplitude MHD linear waves is the fact that nobody so far has proved that there is a link with the well known energy release processes  during explosive solar events (e.g.\ magnetic reconnection in the spontaneously formed current sheets). Petrosian \cite{Petrosian16} and others used the cartoon of the ``standard'' flare model (see Fig. \ref{petro})  to  link the reconnecting current sheet with the weak turbulence needed to accelerate the particles.  As we will see below, fully developed turbulence will naturally provide a link  between energy  release processes and  particle acceleration during impulsive events.

	\subsection{Strong turbulence}
The highly twisted magnetic topologies born out from the current simulations mentioned in the previous section,  lead to a different regime of turbulence with remarkably more complex mixing of unstable current sheets and large amplitude magnetic disturbances; this state of turbulence is also called ``turbulent reconnection'' (see the recent review \cite{Vlahos18} and references therein).

We use the term ``turbulent reconnection" to define an environment were large scale magnetic discontinuities with $\delta B/B >1$ coexist with randomly distributed Unstable Current Sheets (UCS) \cite{Matthaeus86, Lazarian99}.  The importance of turbulent reconnection in many space and astrophysical systems has been discussed in detail  in many recent reviews \cite{Lazarian15, Matthaeus15}.

Isliker et al. \cite{Isliker17a}  consider a strongly turbulent environment as it naturally results from the nonlinear evolution of the MHD equations, in a similar approach as in Dmitruk et al. \cite{Dmitruk04}. Thus, they did not set up a specific geometry of a reconnection environment or prescribe a collection of waves \cite{Arzner04} as turbulence model, but allow the MHD equations themselves to build
naturally correlated field structures (which are turbulent, not random) and
coherent regions of intense current densities (current filaments or CS).

The 3D, resistive, compressible and normalized MHD equations used  are

\beq
\p_t \rho = -\nabla \cdot \mathbf{p}
\eeq
\beq
\p_t \mathbf{p} =
- \mathbf{\nabla}  \cdot
\left( \mathbf{p} \mathbf{u} - \mathbf{B} \mathbf{B}\right)
-\nabla P - \nabla B^2/2
\eeq
\beq
\p_t \mathbf{B} =
-  \nabla \times \mathbf{E}
\eeq
\beq
\p_t (S\rho) = -\mathbf{\nabla} \cdot \left[S\rho \mathbf{u}\right]
\eeq
with $\rho$ the density, $\mathbf{p}$ the momentum density,
$\mathbf{u} = \mathbf{p}/\rho$,
$P$ the thermal pressure,
$\mathbf{B}$ the magnetic field,
\beq \mathbf{E}   = -  \mathbf{u}\times \mathbf{B} + \eta \mathbf{J}\eeq
the electric field,
$\mathbf{J} =  \mathbf{\nabla}\times\mathbf{B}$
the current density, $\eta$ the resistivity,
$S=P/\rho^\Gamma$ the entropy,
and $\Gamma=5/3$ the adiabatic index.

Isliker et al. \cite{Isliker17a} solved the 3D MHD equations numerically
(with the pseudo-spectral method \cite{Boyd2001}, combined with the strong-stability-preserving Runge Kutta scheme \cite{Gottlirb98}) in Cartesian coordinates and by applying periodic boundary conditions to a grid of size $128\times 128\times 128$. As initial conditions they use a fluctuating magnetic field $\mathbf{b}$ that consists of a superposition of Alfv\'en waves, with a Kolmogorov type spectrum in Fourier space, together with
a constant background magnetic field $\mathbf B_0$ in the
$z$-direction, so the total magnetic field is $\mathbf{B}=\mathbf{B}_0 +\mathbf{b}(x,y,z,t)$.  The mean value of the initial magnetic perturbation is  $<b> = 0.6 B_0$, its standard deviation is $0.3 B_0$, and the maximum equals $2B_0$, so that they indeed consider strong turbulence. The initial velocity field is 0, and the initial pressure and energy are constant.

\begin{figure}[h]
	\centering
	\includegraphics[width=0.65\columnwidth]{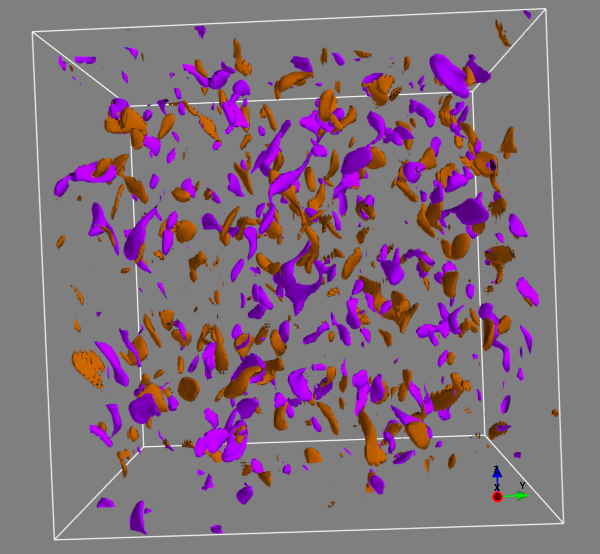}
	\caption{Iso-contours of the supercritical current density component $J_z$ (positive in brown negative in violet).  \cite{Isliker17a}.
	}
	\label{snapshot}
\end{figure}

The structure of
the $z$-component of the current density $J_z$ is shown in Fig.\ \ref{snapshot}.
For the MHD turbulent environment to build, Isliker et al \cite{Isliker17a} let the MHD equations evolve 
until the largest velocity component starts to exceed twice the Alvf\`en speed.
The magnetic Reynolds number at final time is $<|\mathbf{u}|>l/\eta = 3.5\times 10^3$, with $l\approx 0.01$ a typical eddy size.
The overall picture in Fig.\ \ref{snapshot} demonstrates the spontaneous formation of current sheets. This result resembles the 2D simulations of Biskamp and Walter \cite{Biskamp89} almost thirty years ago.  Similar results were obtained by Arzner et al. \cite{Arzner04,   Arzner06}, using Gaussian fields or the large eddy simulation scheme.

\begin{figure}[h!]
	\centering
	\includegraphics[width=0.45\columnwidth]{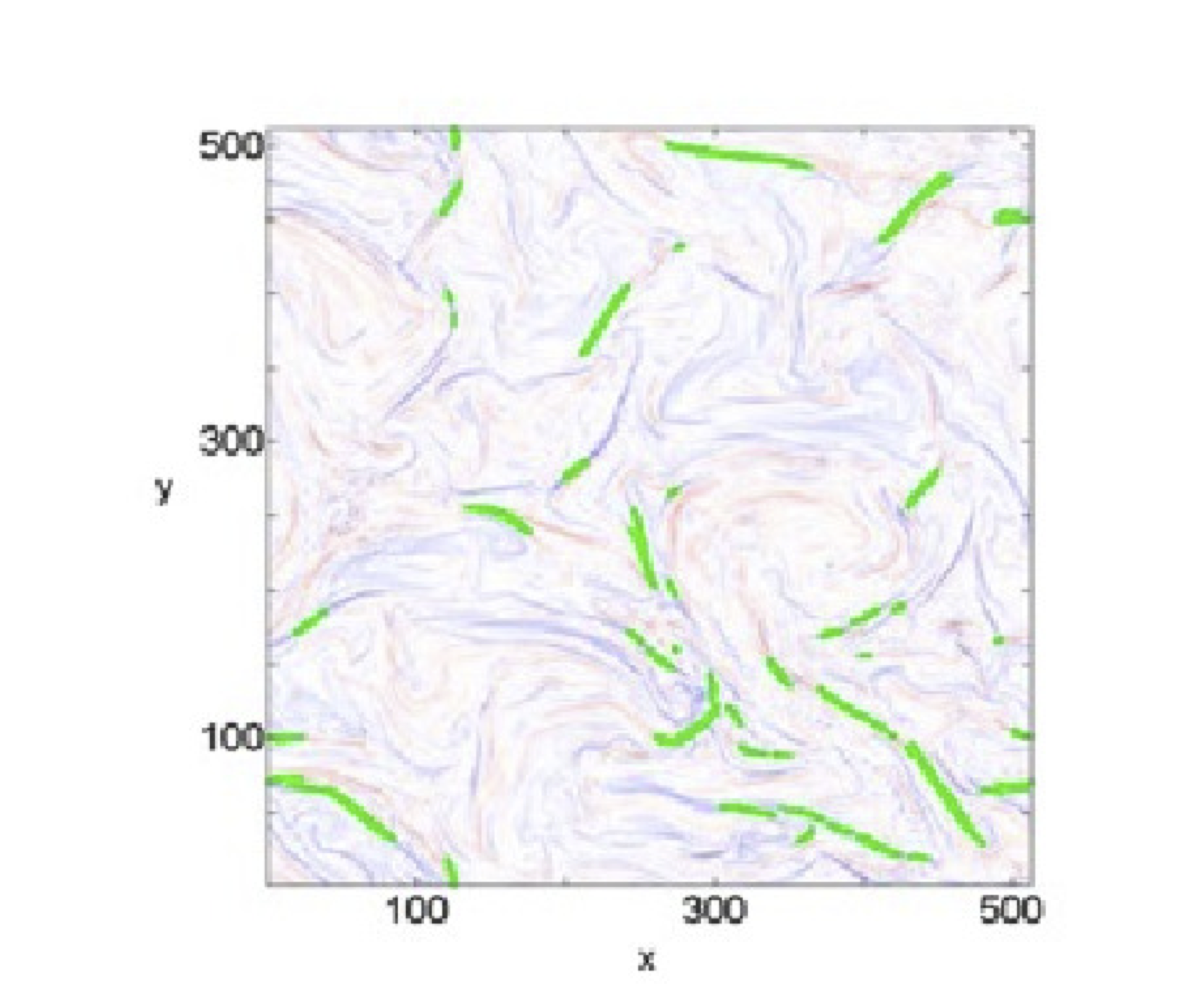}
	\includegraphics[width=0.50\columnwidth]{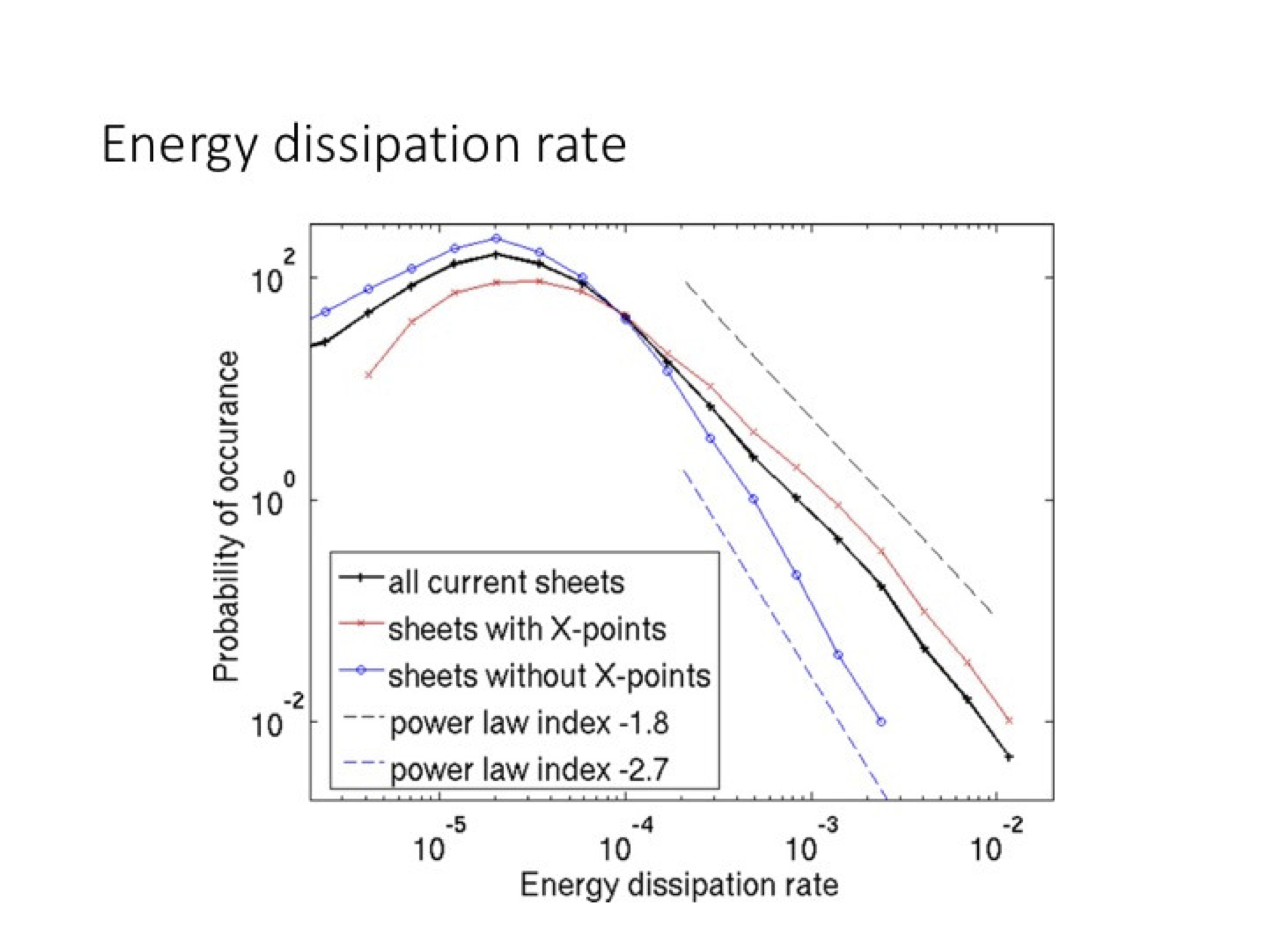}
	\caption{Current density in an $x$-$y$-plane cross section of data. Red indicates negative current and blue indicates positive current. Identified current sheets in the plane are marked by green color.  (b) Probability distribution of the current sheet Ohmic dissipation rate. The distribution from all current sheets (black) shows a power law tail with index near $-1.8$. (From \cite{Zhdankin13}.) }
	\label{Distrcs}
\end{figure}

The statistical properties of the current sheets formed inside strongly turbulent environments
have been analyzed in depth in 2D and 3D simulations by many researchers \cite{Servidio09, Servidio10,  Servidio11, Uritsky10,   Zhdankin13}. Zhdankin et al. \cite{Zhdankin13} developed a framework for studying the statistical properties of current sheets formed inside a magnetized plasma using a 3D reduced MHD code. The current fragmentation in an $x$-$y$-plane, which includes current sheets, is shown in Fig. \ref{Distrcs}. They were able to show that a large number of current sheets do not contain reconnection sites, and likewise, many reconnection sites do not reside inside current sheets.
The most striking characteristic of the current sheets formed spontaneously inside the strongly turbulent plasma is the probability distribution of the dissipated energy $\varepsilon =\int \eta j^2 dV$, which follows a power-law in shape, as reported by Zhdankin et al. \cite{Zhdankin13}
(see Fig.\ \ref{Distrcs}).

Ambrosiano et al. \cite{Ambrosiano88} were the first to analyse the evolution of test particles inside turbulent reconnection modeled by the simulations of Matthaeus and Lamkin \cite{Matthaeus86}. Many years later several researchers returned to this problem and followed the evolution of a distribution of particles inside a snapshot of the 3D evolution of a spectrum of MHD waves \cite{Dmitruk03,  Dmitruk04, Arzner06}.
Isliker et al. \cite{Isliker17a}
use the simulations already reported to
explore the evolution of test particles inside a large scale,
non-periodic,
turbulent reconnection environment. The test-particles are tracked
in a fixed snapshot of the MHD evolution,
		and they evolve the particles
for short times, so they do not probe the scattering of particles off waves, but the interaction with electric fields.
In this particular numerical experiment, anomalous resistivity effects were taken into account.
Physical units are  introduced by using the parameters $L=10^5\,$m for the box-size, $v_A=2\times10^6\,$m/s
for the Alfv\'en speed, and $B_0=0.01\,$T for the background magnetic field.
Isliker et al.\  apply a cubic interpolation of the fields at the grid-points to the actual particle positions.
The relativistic guiding center equations (without collisions) are used
for the evolution of the position $\mb{r}$ and  the parallel component $u_{||}$ of the relativistic 4-velocity of the particles
The test-particles they consider throughout are electrons.
Initially, all particles are located at random positions, they obey a
Maxwellian distribution
$n(W, t=0)$
with temperature $T=100\,$eV. The simulation box is open, the particles can escape from it when they reach any of its boundaries.
\begin{figure}[h]
\centering
\includegraphics[width=0.40\columnwidth]{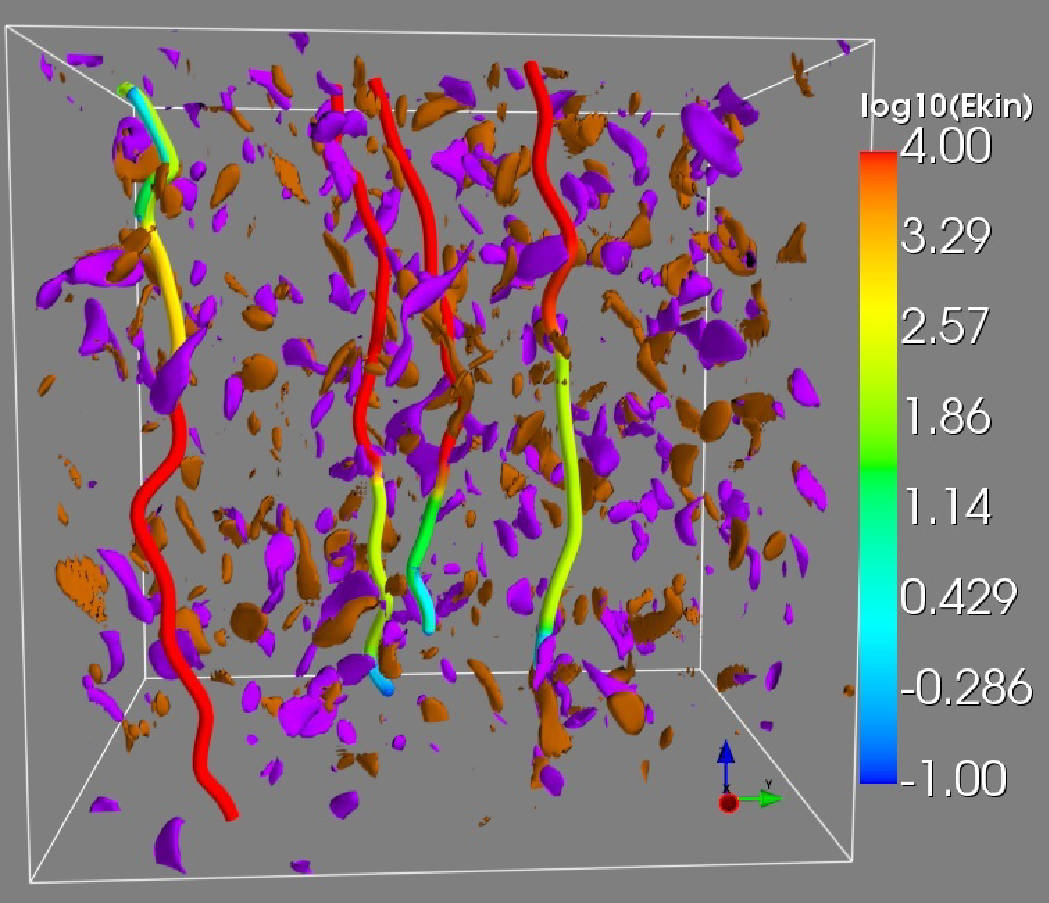}
\includegraphics[width=0.50\columnwidth]{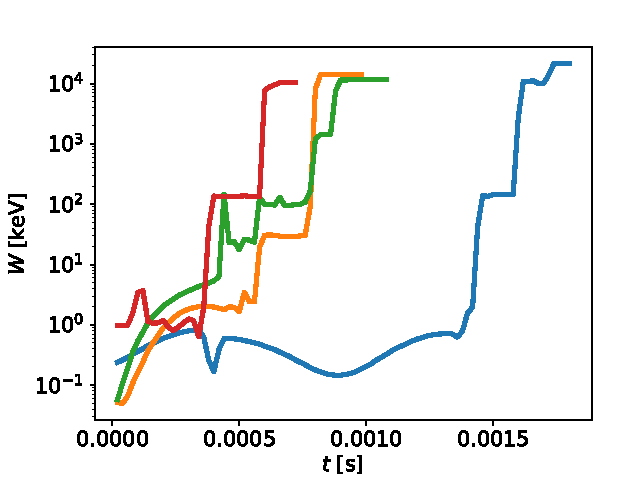}\\\includegraphics[width=0.80\columnwidth]{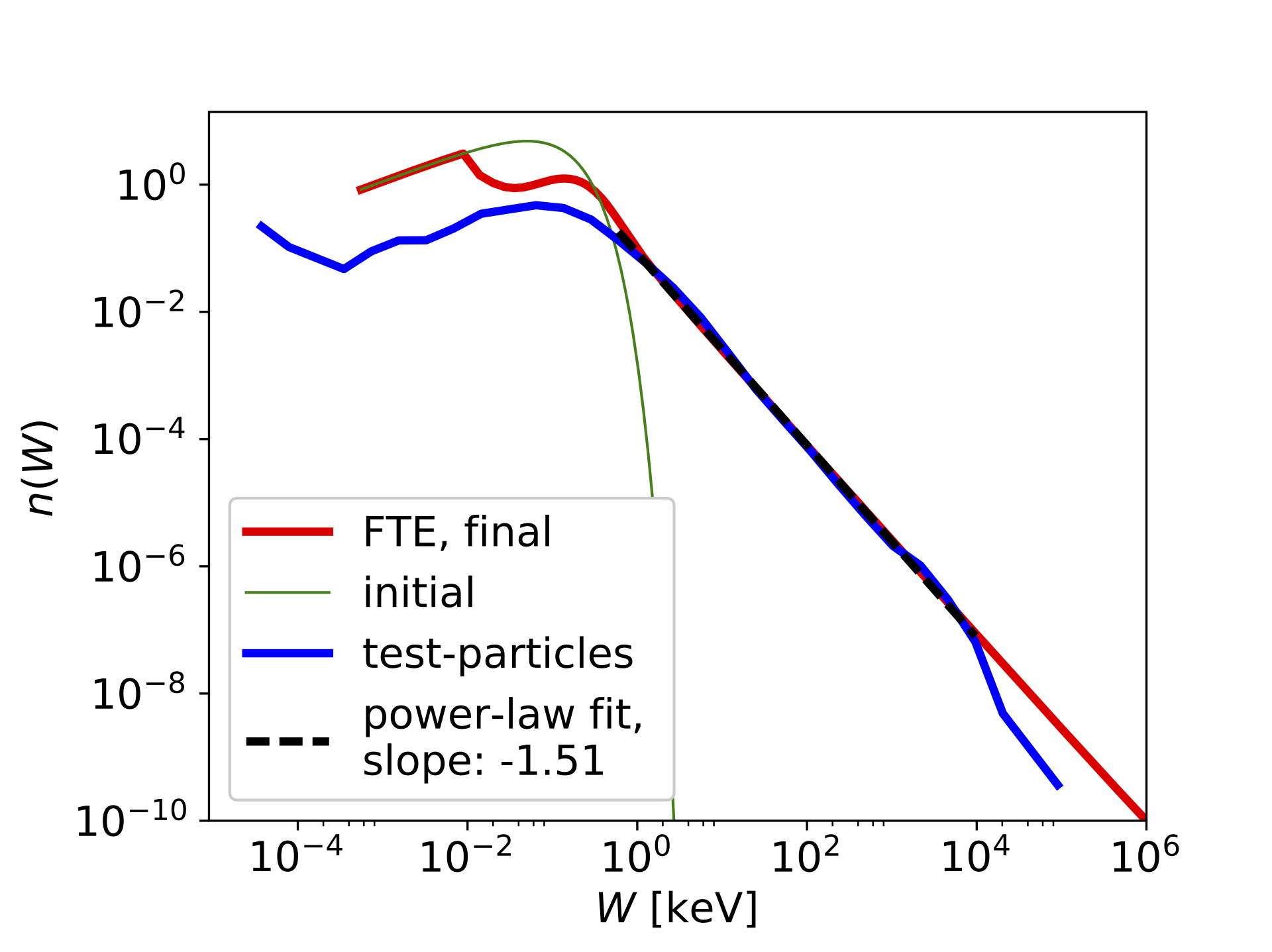}

\caption{ (a) Particle orbits inside the simulation box, colored according to their kinetic energy (b) Typical particle trajectories in energy of some accelerated particles.  (c) Initial and final (at $t
			= 0.002\,$sec) kinetic energy distribution from the test-particle simulations, together with a power-law fit, and the solution of the fractional transport equation (FTE) at final time \cite{Isliker17a}}
\label{Ekin}
\end{figure}

The acceleration process, is very efficient, and they consider a final time
of $0.002\,$s ($7\times 10^5$ gyration periods), at which the asymptotic state has already been reached.
As Fig.\ \ref{snapshot}, Fig.\ \ref{Ekin}a shows the component $J_z$ in the regions of above-critical current density, which clearly are fragmented into a large number of small-scale
current filaments (current-sheets) that represent
coherent structures within the nonlinear, super-Alfv\'enic MHD environment.
The figure also shows the orbits of some energetic particles.
The particles can lose energy, yet they mostly gain energy in a number of sudden jumps in energy (see also Fig.~(\ref{Ekin}b)), the energization process thus is localized and there is multiple energization at different current filaments. Fig.~(\ref{Ekin}c) shows the energy distribution at final time, which
exhibits a clear power law part in the intermediate to high
energy range with power-law index $-1.51$, with a slight turnover at the highest energies.
There is also moderate heating, the initial temperature has roughly been
doubled
(qualitatively similar characteristics of the acceleration process
have been observed in \cite{Arzner04} and in the PIC simulations of
 \cite{Dahlin15,Guo15}).

As shown in \cite{Isliker17a}, the distribution of energy increments
exhibits a power-law tail,
which implies that the particle dynamics is anomalous, with occasionally large energy steps being  made, the particles perform Levy-flights in energy space when their dynamic is interpreted as a random walk.
Isliker et al.  \cite{Isliker17a} introduced a formalism for a fractional transport equation (FTE) that is able to cope with this kind of non-classical dynamics. The solution of the FTE at final time is also shown in
Fig.\ \ref{Ekin}c, obviously the FTE reproduces very well the power-law tail in its entire extent, which confirms that transport in energy space is of fractional nature.
\begin{figure}[!ht]
\centering
\includegraphics[width=0.4\textwidth]{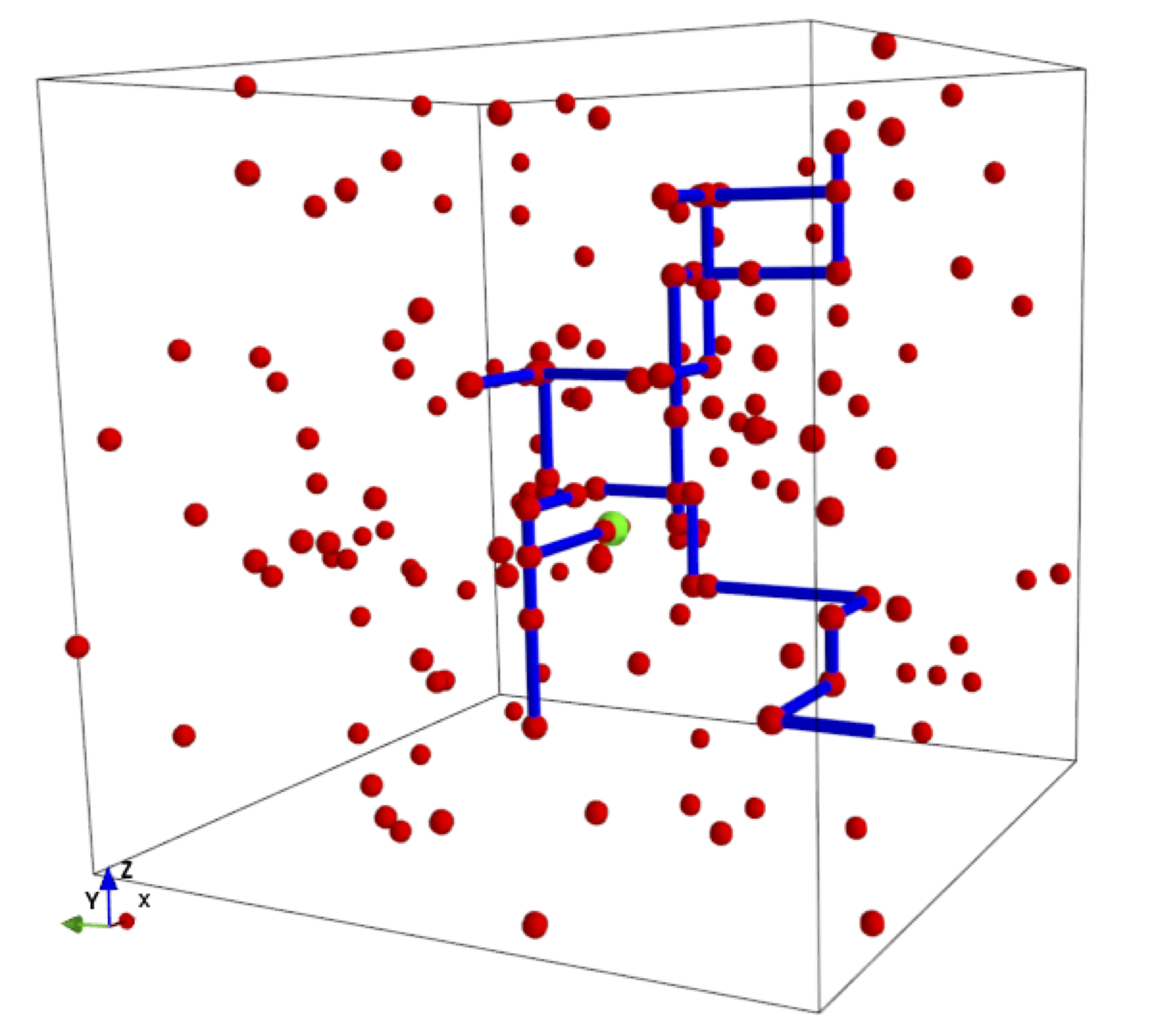}
\includegraphics[width=0.4\textwidth]{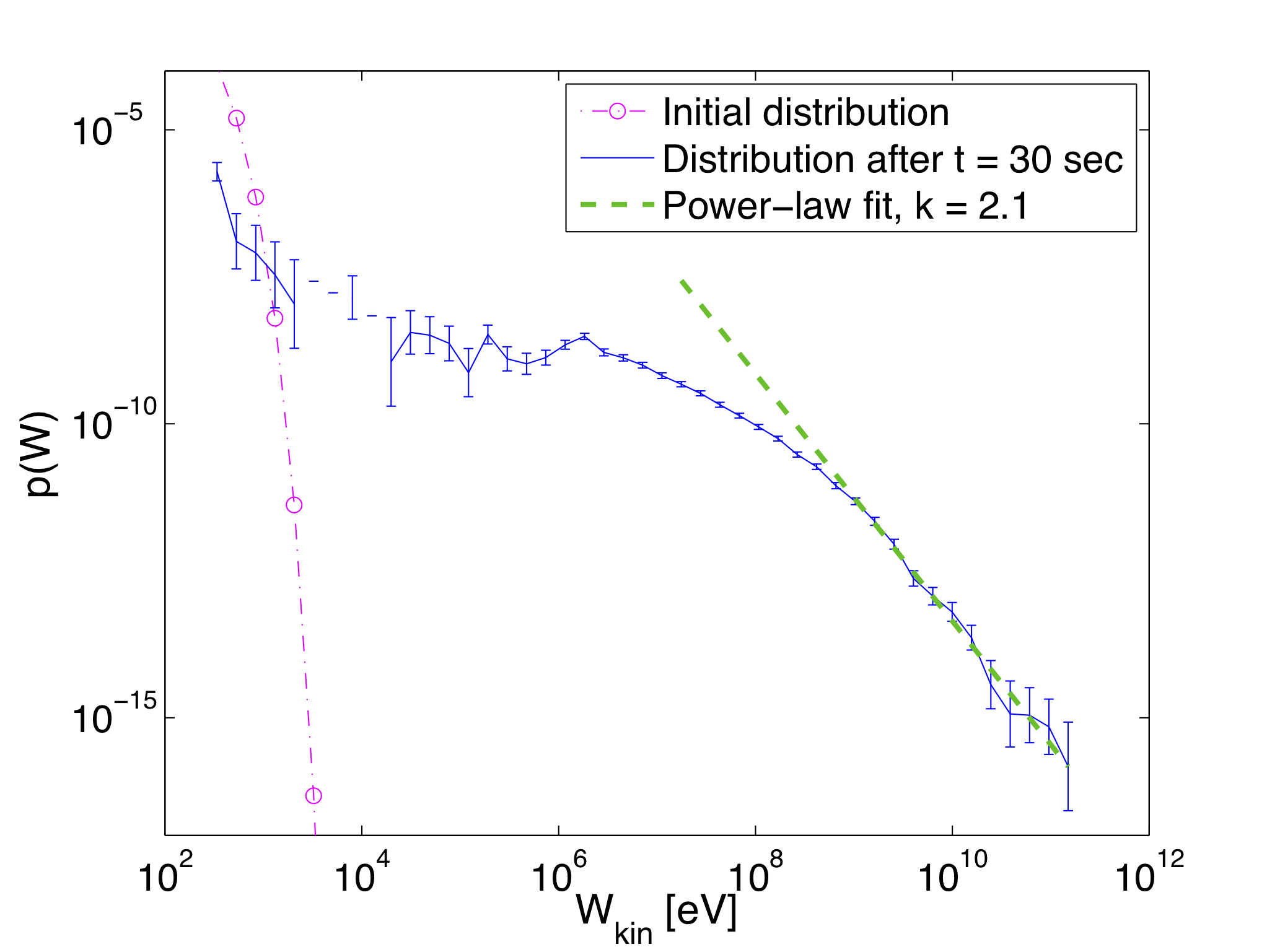}
\caption{ (a) The trajectory of a typical particle (blue tube) inside a grid with linear dimension $L$. Active points are marked by spheres in red color. The particle starts at a random grid-point (green sphere), moves along a straight path on the grid till it meets an active point and then it moves into a new random direction, and so on, until it exits the simulation box. (b) Energy distribution of the ions at $t=0 s$ and $t=30 s$ (stabilised) \cite{Pisokas16}}
\label{Fermi1}
\end{figure}

Pisokas et al. \cite{Pisokas16} analysed the stochastic Fermi acceleration  of ions and electrons interacting with large scale magnetic fluctuations using a simple model, illustrated in Fig. \ref{Fermi1}a. A 3D grid with linear size $L$ is used, and it is assumed that a small percentage of the grid points is active (magnetic disturbances) and the rest is passive. Ions interacting with active grid points gain or lose energy stochastically, following the initial suggestion by Fermi \cite{Fermi49}. The asymptotic energy distribution of the accelerated ions is shown in Fig. \ref{Fermi1}b for parameters similar to the ones in the low corona.

 In turbulent reconnection, stochastic scattering at large scale disturbances co-exists with acceleration at UCSs. It is natural to ask how the ambient particles  react if the two Fermi accelerators act simultaneously. Pisokas et al.\  \cite{Pisokas18} discuss the synergy of the energization at large scale magnetic disturbances  (stochastic scatterers) with the systematic acceleration by UCSs.

\section{Energisation of particles during magnetic flux emergence}

In section 3, we discussed the MHD evolution of data driven large scale magnetic eruptions. One of the important results there was the spontaneous formations of current sheets of different scales at different locations inside the evolving magnetic topology. The evolution of the large scale current sheets and their role in reorganizing the magnetic topology and accelerating particles was not discussed in section 3,  since all these physical processes are below the resolution of the large scale simulations. In this section, 
 we consider higher resolution MHD simulations of
emerging magnetic flux interacting with  the ambient magnetic field and forming  a large scale current sheet that fragments and becomes an efficient particle accelerator.

Emerging magnetic flux into pre-existing magnetic fields drives the formation of large scale reconnecting current sheets in tens of minutes and can be the source of several eruptive phenomena, e.g. flares, prominence eruptions, jets, CMEs \cite{Heyvaerts77, Archontis04, Galsgaard05,Archontis04, Archontis05, Archontis12a, Archontis12b, Archontis13,Moreno-Insertis13,Karibabadi2013c,Jiang16,  Raouafi16, Wyper16, Wyper17}. The emerging flux will drive a standard or a blowout jet, which can be the source of impulsive or gradual SEP events (see the recent review in Ref.\ \cite{Raouafi16}).

\begin{figure}[!ht]
\centering
\includegraphics[width=0.8\textwidth]{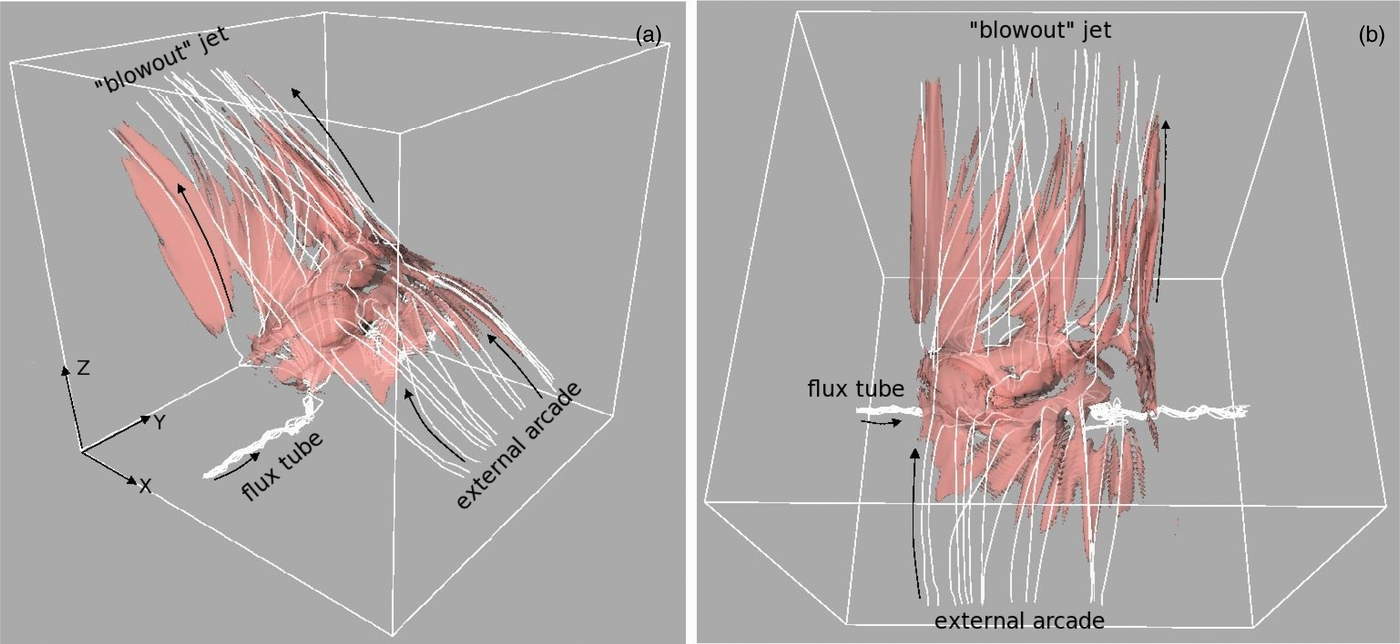}
\caption{Side-view (panel (a)) and top-view (panel (b)) of the 3D fieldline topology and velocity (isosurface $\geq 200 km s^{-1}$) during the blowout jet emission ($ t=54\,minutes $). The direction of the fieldlines is shown by the (black) arrows. \cite{Archontis13}
\label{fl1}}
\end{figure}

Arhontis and Hood \cite{Archontis13} use a 3D resistive MHD code to follow the emergence of new magnetic flux into the pre-existing magnetic field in the corona. The formation and subsequent fragmentation of the large scale reconnecting current sheets is obvious in their numerical study (see Fig. \ref{fl1}). Details of their simulation and their results can be found in their article.

\begin{figure}[!ht]
\centering
\includegraphics[width=0.3\textwidth]{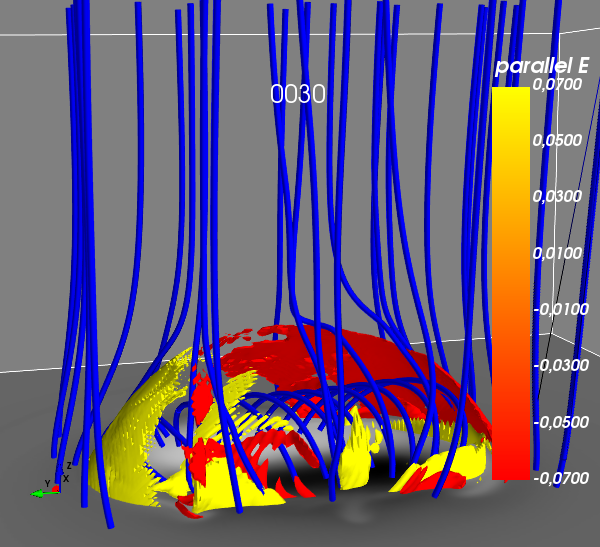}
\includegraphics[width=0.3\textwidth]{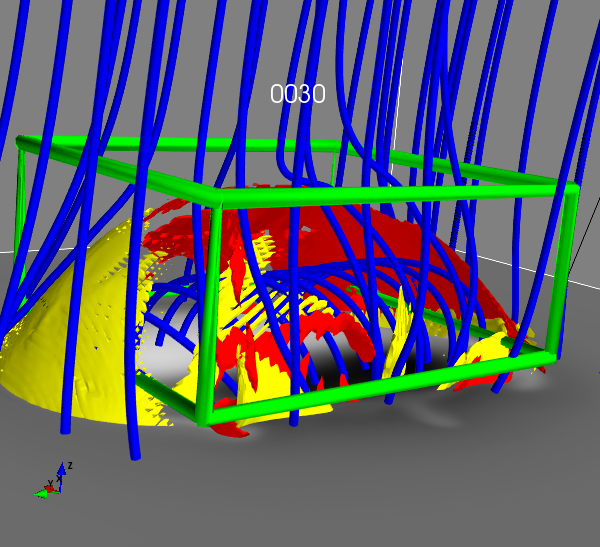}
\caption{MHD simulations, zoom into the coronal part:
Left: Magnetic field lines (blue tubes),
together with an iso-contour plot 
of the parallel electric
field (red and yellow 3D-surfaces).
At the bottom $x$-$y$-plane,
the photo-spheric component $B_z$ is shown as a 2D filled contour plot.
Right: As left, zoomed, and the
region in which the spatial initial conditions for the particles are chosen is
out-lined by a green cube \cite{Isliker19}. \label{fl2}}
\end{figure}

Isliker et al. \cite{Isliker19} use the results from the numerical simulations of Archontis and Hood \cite{Archontis13} but focus on the statistical properties of the electric fields and the energy transport of electrons in the vicinity of the fragmented large scale current sheet (see Fig. \ref{fl2}).
They first consider the energization of particles at the standard jet, snapshot 30.
Electrons are considered as test-particles, and
the standard integration time is $0.1\,$sec,
and 100'000 particles are traced, in any case by using
the relativistic guiding center approximation to the equations of motion.
The initial spatial position is uniform random in the region around
the main reconnection region, as out-lined by the green cube
in Fig.\ \ref{fl2}. The initial velocity is random with Maxwellian
distribution (i.e.\ Gaussian distribution of the velocity components),
with temperature $\approx 9 \times 10^5\,$K.
For each simulation, a set of 100 monitoring times has been predefined,
including the final time, at which the velocities and positions of the particles are
monitored for the purpose of a statistical analysis to be done at equal
times for all the particles. Separate track is kept of the particles
that leave before the final time.
\begin{figure}[!ht]
\centering
	\includegraphics[width=0.7\textwidth]{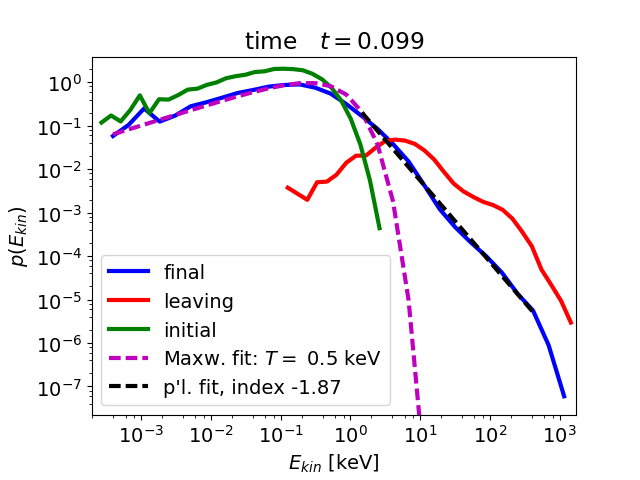}
	\caption{
		Snapshot 30: Kinetic energy distribution of electrons after $\approx 0.1\,$s,
		without
		collisions, together with a fit at the low-energy, Maxwellian part and the
		high energy, power-law part, the initial distribution,
		and the
		distribution of the leaving particles (for every particle at the time it
		leaves).
		\cite{Isliker19} \label{Ekin1}}
\end{figure}

Fig.\ \ref{Ekin1} shows the distribution of the kinetic energies
of the particles after $0.1\,$s, together with the initial distribution
and the distribution of the leaving particles (as collected at the times
the individual particles leave). The final energy distribution is of
Maxwellian shape at the low energies, and exhibits a slightly modulated
power-law tail.
The maximum energy reached is about $1\,$MeV, and
a power-law fit to the tail of the kinetic energy distribution
yields an index of about -1.87.
13\% of the 100'000 particle that are traced have left after $0.1\,$s,
and they
have energies in the same range than those
that stay inside, with a modulated power-law tail
that is
steeper though, with index -2.98 at the highest energies (the fit is not shown).

The energy distribution of the leaving particles
shows a functional
form at low energies (between $0.1$ and $10\,$keV) that is reminiscent of a Maxwellian, and
a respective fit reveals a temperature of about $13.3\,$keV
(see Fig.\ \ref{Ekin1}, the fit itself is not shown). Although the statistics is not
very good, we can interpret these particles as belonging to a super-hot
population. It is to note though that the energies are monitored
at different times for each particle, so the distribution is asynchronous.

For the particles that stay inside,
the Maxwellian shape of the energy distribution is well
preserved at low energies,
and there is heating from the initial $0.24\,$keV to $0.50\,$keV after
$0.1\,$s,
as the Maxwellian fit in Fig.\ \ref{Ekin1}, reveals.

As shown in \cite{Isliker19}, the distribution of energy increments
	has a power-law tail, as in the case of strong turbulence reported above (\cite{Isliker17a}),
	so that also in the case of emerging flux, the transport in energy space is of fractional nature.

\section{Acceleration of particles by CME driven shocks}%

One of the prominent acceleration mechanisms in astrophysics are shock waves. The acceleration of particles in shocks, developed by the eruption of unstable magnetic structures interacting with  the ambient magnetic field inside the solar corona, is still an open problem.  We are still lacking a clear model of the shock driven by a CME near the Sun, even if shock acceleration near the Sun appears to be the most promising acceleration mechanism for SEPs \cite{Lee12}.

The theory of shock acceleration has been developed in relatively simple magnetic topologies of planar shocks. The angle between the direction of the ambient magnetic field upstream of the shock with the velocity of propagation of the shock is an important parameter for the acceleration process. When the shock is propagating along the direction of the magnetic field (parallel shock),
the acceleration is due to the trapping of particles around the shock discontinuity by weak turbulence upstream and downstream. This mechanism is called Diffusive Shock Acceleration (DSA) and has been analysed extensively in the current  literature \cite{Burgess13, Caprioli14a,Caprioli14b, Caprioli14c}.

\begin{figure}[!ht]
\centering
\includegraphics[width=0.40\textwidth]{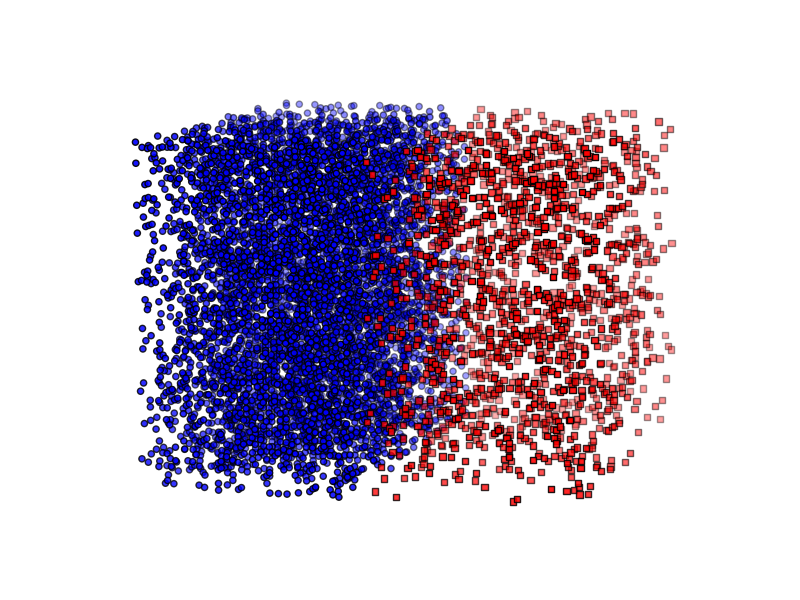}
\includegraphics[width=0.55\textwidth]{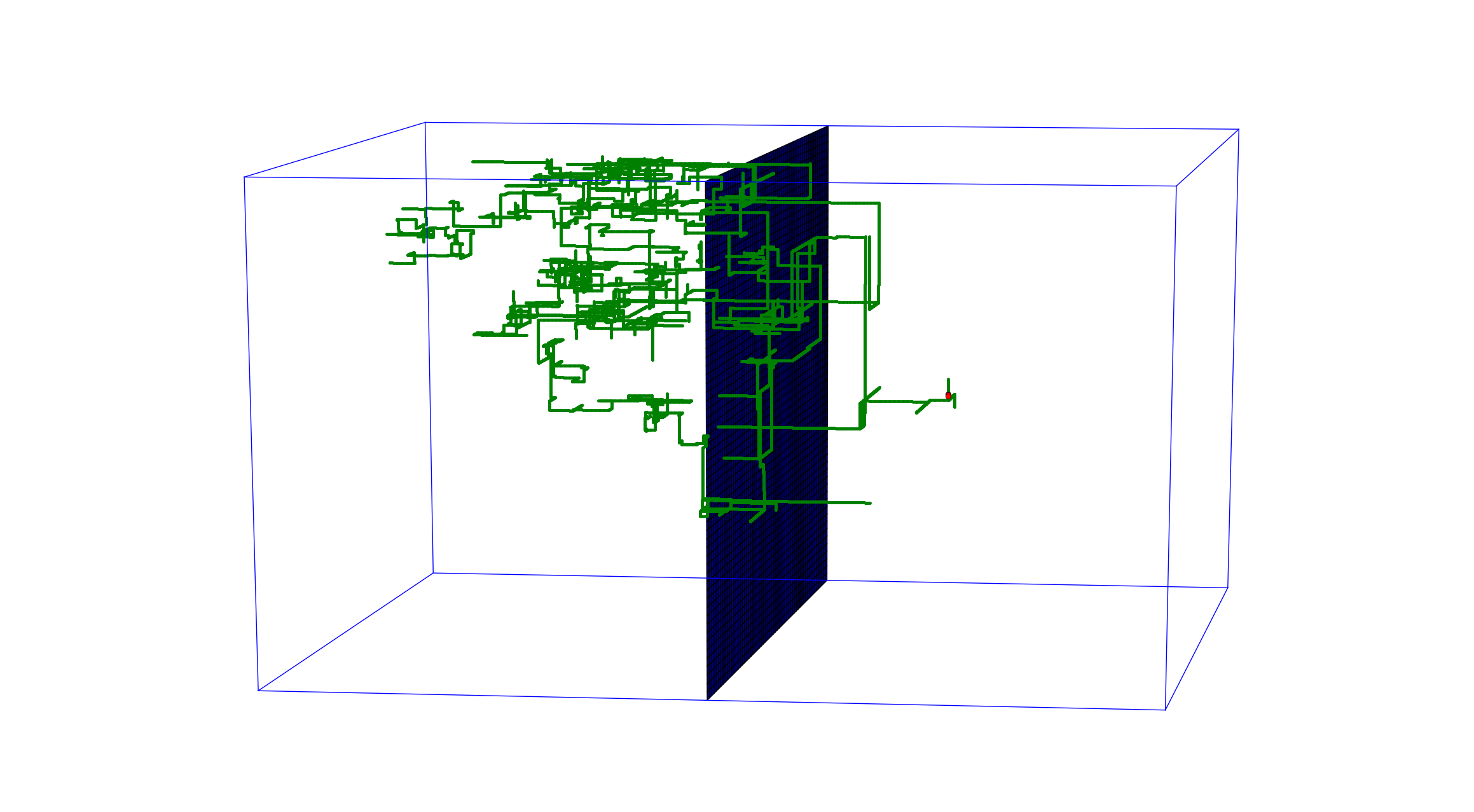}
\caption{(a) Small version of the simulation box with the planar shock wave in the middle. The strongly turbulent environments upstream and downstream are acting as active scatterers. Particles not only return back to the shock, as in the case of the traditional DSA model, but also gain energy upstream and downstream. (b) Trajectory of a typical particle inside the simulation box.
\cite{Garrel18} . \label{fsck}}
\end{figure}

The mechanism of particle acceleration  by shocks propagating perpendicular to the upstream magnetic field is different and relies mainly on the direct acceleration of particles drifting along the convective electric field $\vec{E}=-\vec{U} \times \vec{B}$ \cite{Holman1985, Sandroos06}.  It is obvious that such a clear division of the two processes in realistic shocks traveling inside the solar corona is impossible, so the mixing of the two in a strongly turbulent plasma is more relevant and important.

Recently, a departure from the traditional approach was made by assuming that the turbulence upstream and downstream can reach very high amplitudes, $(\Delta B/B)>>1 $,  and turbulent reconnection will set in \cite{Zank15, leRoux16}. This is true especially when a CME and the shock are propagating against a preexisting turbulent magnetic field or ambient magnetic structures, like the termination shock \cite{Burgess16}.  Yang et al. \cite{Yang16}  simulated the interaction of the turbulent solar wind with the Earth's magnetic field using 3D Particle In Cell simulations.
The CME driven shocks have many similarities with the Earth's Bow Shock.
Garrel et al. \cite{Garrel18}   discuss the flowing question:  if large amplitude magnetic disturbances are present upstream and downstream of a shock   then Turbulent Reconnection  will set  in and will participate not only in the elastic scattering of particles but also in their heating and acceleration (see also \cite{Afanasiev14}). Garrel et al.\ demonstrate that large amplitude magnetic disturbances and UCS, spontaneously formed in the strong turbulence in the vicinity  of a shock, can accelerate particles as efficiently as DSA in large scale systems and on long time scales (Fig. \ref{fsck}).   They show that the asymptotic energy distribution of  particles accelerated by the combined action of DSA and Turbulent Reconnection has very similar characteristics with the one due to DSA alone, but the synergy of DSA with Turbulent Reconnection is much  more efficient: The acceleration time is an order of magnitude  shorter and the maximum energy reached two orders of magnitude higher.  They claim that DSA is the dominant acceleration mechanism in a short period before Turbulent Reconnection is established, and then  strong turbulence will dominate the heating and acceleration of the particles. In other words, the eruptive large scale magnetic structure and the shock formed ahead of a CME  act as the mechanism to set up a strongly turbulent  environment, in which the acceleration mechanism will ultimately be the synergy of DSA and Turbulent Reconnection.

\section{Discussion} 
We can now pose an important question: If the standard flare cartoon is not a realistic representation of the physical processes related to solar eruptions, what will be its alternative?

The data driven approach presented in section 3 is much closer to a realistic representation of how solar eruptions spontaneously form current sheets in different parts of the stressed and twisted  large scale magnetic topology. The evolution of the large scale current sheets presented in section 5 shows how a fragmenting current sheet drives strong turbulence locally.  The distinction of strong and weak turbulence and their efficiency in accelerating electrons and ions was analysed in section 4. Combining the findings of these  sections we can redefine the term "flare", as appearing in several articles studying impulsive SEP events, with the phrase "impulsive energy release by the spontaneous formation of current sheets that fragment  and form a strongly turbulent environment locally".

In section 6 we analyse the heating and acceleration of particles in shocks formed ahead of the CME and propagating inside the solar wind, which is always in the sate of strong  turbulence.  In many ways the CME driven shock is similar to the Earth's Bow shock and departs  from the simple model of the Diffusive Shock Acceleration since the scattering of the particles upstream and downstream accelerates the particles as well. 

Flare and CME driven shocks  rely on  strong  turbulence in order to heat and accelerate the plasma particles.  The driver of the strong turbulence in the case of a "flare" is the large-scale current sheet formed spontaneously in the low corona and fragmenting impulsively, and in the case of CME driven shocks the driver  is the solar wind.

Let us now return to the key observational points mentioned in Section 2  and try to connect them with the theoretical arguments presented above.

\begin{enumerate}
	\item The impulsive-flare related SEP events are associated with localized sources close to the Sun. Gradual-CME related SEP events are usually determined at widely separated locations in the heliosphere.
	
	\item SEP events with characteristics resembling impulsive events can also be detected over a wide longitudinal range.
	
	\item The simplistic dichotomy of the SEP events into impulsive and gradual is not present in all SEP events.

	\bigskip
	
		{\it The model presented here supports the view  that the spontaneous formation of current sheets with different scales will take place  in different parts of the large scale erupting structure.  Therefore the impulsive injection of particles from many localized volumes, which are widely separated in space, should be expected.  The CME driven  shock will be present in most eruptions and can also efficiently accelerate ions and electrons,  as efficiently as a ``flare'', as we have shown in section 6. So the dichotomy  between impulsive and gradual events depends mainly  on the magnetic topology  and the transport properties of the particles inside the complex and turbulent magnetic structures. In the erupting structure, the strong turbulence driven by the fragmented current sheet (impulsive SEP events) and the shock (gradual SEP events) will coexist. }
		
		\bigskip

		\item Observations are consistent with the acceleration process occurring  over a wide range of longitudes rather than a small source region, and the accelerated particles could be transported before being injected at distant longitudes.

	{\it This is an important point and should be analysed in future studies. How the strongly turbulent plasma and the stochastic field lines generated by the current fragmentation will influence the anomalous transport of particles from the low corona to the interplanetary space remains an open problem.}
	
	\bigskip
	
	\item The comparison between CME or shock parameters and SEP properties show significant correlations, better than the correlations with flare parameters.
	
	\bigskip

	{\it The eruptive magnetic structures forming a shock ahead of the CME will transport the accelerated particles in the interplanetary space much more easily than the impulsive events hidden inside the complex magnetic topology and forming current sheets. }
	
	\bigskip
	
	\item SEP events, either gradual or impulsive, were found to have high association with both Type III and Type II radio bursts.
	
	\bigskip
	
	 {\it Strong turbulence  accelerates very efficiently both  electrons and ions. The only difference is in the acceleration time, the acceleration time for ions is ten times longer. We then expect  that type III radio bursts will always be injected at the strongly turbulent locations associated with the fragmented current sheet.  The type II radio burst will always be present when shocks are formed ahead of the CME. The fact that we do not always detect them is mainly due to observational limitations. }
	
\bigskip
	
\item Gradual events are in generally  considered  to have a composition similar to that of the corona or solar wind, while impulsive events typically have enhanced element and isotope ratios.
	
	\bigskip

{\it We did not address this very  important and crucial issue in this review, since the efficiency of strong turbulence in the acceleration of particles with different elemental abundances and isotopes has not been studied yet.} 

\bigskip	
	
\item The energy spectra based on the Ground Level Enhancements show a double power-law, with the break between 2-50 MeV.
	
\bigskip
		
	{\it This observation is not only related with the sources of SEP events. The transport and acceleration of particles in the IP space is crucial and still remains an open problem. In Section 4, 5 and 6 we have reported     the expected power low slopes of the accelerated particles in the coronal sources.}
	
	\bigskip
		
	\item  Observations have revealed that most of the SEP events are associated with both flares and CMEs. Several intense flares that seem not to be accompanied by a CME, 
	also were lacking SEPs.
	
	\bigskip
	
	{\it We claim that these peculiarities are related with the geometrical characteristics of the erupting magnetic topology and the anomalous transport of particles in the IP space.}
	
\end{enumerate}

\section{Summary}
In the  current literature, the sources of SEPs remain confined to simple sketches and many assumptions based on monolithic current sheets, weak turbulence theory, and  shocks traveling in ordered magnetic topologies, forming quasi-parallel or quasi-perpendicular shocks.  The acceleration mechanisms are analyzed separately from the evolving and dynamic magnetic structures in an eruptive magnetic topology.

In this review, we propose a different approach based on the recent developments in the study of eruptive magnetic structures and the analysis of particle acceleration in strongly turbulent plasmas. We claim that we cannot develop a realistic model for the sources of SEP events if the eruptive 3D magnetic topology remains on the level of simple sketches.

We suggest that the recent 3D data driven studies of eruptive phenomena, based on the initialization of resistive 3D MHD codes with Non-Linear Force Free Extrapolations of the observed magnetograms of specific active regions, driven by the turbulent photospheric activity \cite{Jiang16, Inoue18, Amari18}, is clearly closer to  reality.

It is clear that an eruptive realistic topology naturally forms reconnecting current sheets on all scales, and the large scale reconnecting current sheets fragment, generating a distribution  of strongly turbulent locations along the 3D structure.  The fragmented large scale current sheets and the interaction of the unstable magnetic structures form a turbulent reconnecting environment (a mixture of large scale magnetic fluctuations and reconnecting  current sheets)
along the erupting magnetic structure. Emerging magnetic flux will also initially form  large scale reconnecting current sheets, which will fragment, forming a strongly turbulent environment and large scale jets.

The CME-driven shock follows the same evolution as the main body of the eruptive structure. The presence of turbulent reconnection ahead (solar wind) and behind the shock  play a key role in the long lasting acceleration of ions and electrons.  Karimabaldi et al.  \cite{Karimabadi2014, Yang16}  proposed   that the shock is linked with turbulence and reconnecting current sheets  in a strongly turbulent environment.

How particles are accelerated in a strongly turbulent environment is a new topic that is
under development in the current literature \cite{Vlahos18}. The main characteristic of the interaction  of ions and electrons with a  strongly turbulent plasma is the intense heating and the formation of power law tails. The energy transport properties of the energized particles inside a strongly turbulent environment differ radically from the standard Fokker-Planck approach, since the interaction of the particles with strong turbulence is anomalous, and it can be described with a Fractional Transport Equation \cite{Isliker17, Isliker17a}.

We left outside our discussion in this review  an important question: Can strong turbulence explain the abundance of elements and isotopes in SEPs?   We hope that this question will be analysed soon in the context of strongly turbulent acceleration. A key observation, which we hope will be addressed properly with the Solar Probe, is the relative role of strong turbulence vs shock acceleration in the vicinity of the CME-driven shock front, close to the solar corona. Are phenomena present that are similar to the ones observed in the Earth's Bow shock/magnetosphere ?

Closing this review, it is important to stress once again that   there is a close link between the acceleration mechanisms  and the evolving large scale erupting magnetic topology.  The evolving topology is  hosting  the particular acceleration mechanisms through the distributed energy release sites along the large scale structure.

\aucontribute{LV oversaw the writing  of the entire manuscript, and led the writing on sections 1, 3, 4, 5, 7. AA, AK, AP led the writing on section 2, HI the writing of sections 4 and 5. All authors have read and approved the manuscript.}
\competing{The authors declare that they have no competing interests}

\funding{ Part of this study was funded by the European Union (European Social Fund) and the Greek national funds through the Operational Program ``Education and Lifelong Learning'' of the  National Strategic Reference Frame Work Research Funding Program: Thales. Investing in Knowledge society through the European Social Fund.}

\ack{We thank the referee for his/her constructive comments and suggestions. The present work benefitted from discussions held at the International Space Science Institute (ISSI, Bern, Switzerland) within the frame of the international team ``High EneRgy sOlar partiCle (HEROIC)'' led by Dr. A. Papaionnou.}

\end{document}